\documentclass[useAMS,usenatbib]{mn2e}

\usepackage{amssymb}
\usepackage{amsmath}
\usepackage{graphicx} 
\usepackage{epsfig}
\usepackage{color}
\usepackage{bm}
\usepackage{ulem}
\usepackage{color}

\newcommand{\hatvi}{{\bm{{\rm\hat v_i}}}}
\newcommand{\vv}{{\bm{{\rm v}}}}
\newcommand{\vvi}{{\bm{{\rm v_i}}}}
\newcommand{\ei}{{\bm{{\epsilon_i}}}}
\newcommand{\muj}{{\bm{{\rm\mu_j}}}}
\newcommand{\Sigmaj}{{\bm{{\rm\Sigma_j}}}}
\newcommand{\Si}{{\bm{{\rm S_i}}}}
\newcommand{\Tij}{{\bm{{\rm T_{ij}}}}}
\newcommand{\kms}{km s$^{-1}$}
\newcommand{\kmss}{km$^{2}$ s$^{-2}$}

\author[Q. Xia et al.]{Qiran Xia$^{1,2}$\thanks{qiranxia@gmail.com}, Chao Liu$^{1,3}$, Yan Xu$^{1,3}$, Shude Mao$^{1,4}$, Shuang Gao$^{1,5}$, Yonghui Hou$^{6}$, \newauthor Ge Jin$^{7}$,  Yong Zhang$^{6}$\\
$^1$ National Astronomical Observatories, CAS, 20A Datun Road, Chaoyang District, 100012, Beijing, China\\
$^2$ University of Chinese Academy of Sciences, Beijing 100049, China\\
$^3$ Key Laboratory  of Optical Astronomy, National Astronomical Observatories, CAS, 20A Datun Road, Chaoyang District, 100012, Beijing, China\\
$^4$ Jodrell Bank Centre for Astrophysics, The University of Manchester, Alan Turing Building, Manchester M13 9PL, UK\\
$^5$ LAMOST fellow\\
$^6$ Nanjing Institute of Astronomical Optics \& Technology, National Astronomical Observatories, Chinese Academy of Sciences, Nanjing 210042, China\\
$^7$ University of Science and Technology of China, Hefei 230026, China}

\title{The velocity distribution in the solar neighbourhood from the LAMOST pilot survey}

\begin{document}
\include{journaldefs}

\date{Accepted 2014 December 9. Received 2014 December 9; in original form 2014 March 28   }

\pagerange{\pageref{firstpage}--\pageref{lastpage}} \pubyear{2014}

\maketitle

\label{firstpage}

\begin{abstract}
We use about 15,000 F/G nearby dwarf stars selected from the LAMOST pilot survey to map the U-V velocity distribution in the solar neighbourhood. An extreme deconvolution algorithm is applied to reconstruct an empirical multi-Gaussian model. In addition to the well known substructures, e.g., Sirius, Coma Berenices, Hyades-Pleiades over-densities, several new substructures are unveiled. A ripple-like structure from (U, V)=$(-120, -5)$ to $(103, -32)$\,\kms\ is clearly seen in the U-V distribution. This structure seems associated with resonance induced by the Galactic bar, since it is extended in U while having a small dispersion in V at the same time. A ridge structure between (U, V)=$(-60, 40)$ and $(-15, 15)$\,\kms\ is also found. Although similar substructures have been seen in the \textit{Hipparcos} data, their origin is still unclear. Another compact over-density is seen at (U, V)=$(-102, -24)$. With this large data sample, we find that the substructure located at V$\sim-70$\,\kms\ and the Arcturus group are essentially parallel in V, which may indicate that they originate from an unrelaxed disk component  perturbed by the rotating bar.    
\end{abstract}


\section{Introduction}\label{sect:intri}
The origin of the substructures in the velocity distribution in the solar neighbourhood is not clear, although several theories have been proposed. For a long time, it was believed that the substructures are associated with disrupted stellar clusters (\citealt{Kapteyn05, Eggen65, Skuljan97} etc.), which is probably where the name \textit{stellar moving groups} comes from. When \textit{Hipparcos} \citep{Perryman97} data became available, with accurate parallaxes and proper motions for tens of thousands nearby stars, \citet{Dehnen98} recognized that the moving groups follow the asymmetric drift relation, i.e., older groups are hotter and on eccentric orbits, which seems associated with resonance induced by non-axisymmetric force. \citet{Famaey05, Famaey07, Famaey08} found that the member stars of some moving groups have a wide range of age, mass, and metallicity, which does not imply a disrupted cluster origin. \citet{Bensby07} also claimed that the Hercules stream has a range of age and chemical abundances. These observations indicate that stellar cluster disruptions are not responsible for most of the well-known substructures, e.g., the Sirius, Hyades, Pleiades, Hercules streams, but may be for a few moving groups, e.g., HD 1614 \citep{Silva07}. Consequently, theoretical works tend to explain the moving groups as the dynamical effects of the bar and/or spiral arms of the Milky Way.

\citet{Weinberg94} showed that the Galactic bar can lead to distinctive stellar kinematics near the Outer Lindblad resonance (hereafter OLR). \citet{Dehnen00}, who analysed the properties of several types of the closed stellar orbits near to the OLR using test particle simulations, concluded that the unstable $\rm x_{1}^{*}(2)$ orbits produce the valley between the Hercules substructure and the majority of the stars in the U-V distribution, where U is the radial velocity (U$>$0 towards the Galactic centre) and V is the tangential velocity. In his model, there are a few elongated features, one of which, located at U$>0$, V$<$0, is caused by stars on nearly closed orbits with large perturbative amplitudes around the OLR. Moreover, he reproduced ripple-like feature associated with the outer 1:1 resonance in his simulation. Variations in the U-V distribution associated with the angle of the bar, the strength of the bar, the OLR location, and the shape of the rotation curve were also constrained in his work.
\citet{Fux01}, on the other hand, focused more on the orbital analysis and split the stellar orbits into regular, which belong to the disk, and chaotic, which are migrated from the region within the co-rotation radius of the bar. As a consequence, he inferred that the Hercules over-density is due to the outward mixed stars on chaotic orbits. In his test particle simulation, resonances generate distinct arcs in the velocity plane, which  open towards lower angular momentum. 
Unlike \citet{Dehnen00}, the 1:1 resonance plays no role in the Fux model.
On the contrary, the 4:1 resonance feature, which consists of $\rm x_{1}(2)$ orbits, can be recognized, particularly for cases with $R/R_{\rm OLR}<1$ (inside the OLR).
Note that there are no substructures corresponding to the cold moving groups, e.g., Sirius, Hyades, Pleiades, etc., in both the \citet{Dehnen01} and \citet{Fux01} simulations. These smaller scale substructures are more likely related to the local spiral arm(s) \citep{Antoja11, Quillen11}.

With only data from the solar neighbourhood, the origin of the velocity substructures may not be constrained well. Since simulations can easily predict the velocity distribution at different positions in the Galactic disk (\citealt{Dehnen00, Fux01, Bovy09}; etc.), stellar kinematics beyond the solar neighbourhood are required and will play an important role in this study. Recently, \citet{Antoja12} used RAVE data to map the velocity distribution at about 1\,kpc around the Sun and found that the Hercules over-density is a decreasing function of the Galacto-centric radius. Based on a test particles simulation similar to \citet{Dehnen00}, \citet{Antoja13} inferred that this is induced by the rotating Galactic bar. \citet{Liu12} also found that the radial velocity of red clump stars shows a bifurcation at 10-11\,kpc in Galacto-centric radius in the Galactic anti-centre direction, which may again be associated with the bar.

The on-going LAMOST survey \citep{Zhao12} will observe several million dwarf stars in low resolution, and will provide the largest spectroscopic sample within a few kpc around the Sun \citep{Deng12}. This dataset will provide the velocity distribution at different positions in the azimuth-radius plane, and thus will enable the investigation of the role of resonances induced by the rotating bar in the velocity distribution. 

The LAMOST pilot survey has publicly released in excess of 600,000 stellar spectra. We select more than 14,000 F and G dwarf stars from the pilot survey and map them onto the U-V plane. In this paper, we show new evidence, within 500\,pc around the Sun, associated with resonances of the Galactic bar. The structure of the paper is as follows. In section 2, we outline the extreme deconvolution method developed by \citet{Bovy09}, which we apply to derive the intrinsic distribution of the LAMOST stars in the U-V plane. We validate the method using mock data with various velocity errors to help identify the detection limits of the substructure scale. In section 3, we introduce how the stars are selected and their distances estimated. The selection bias of the data is taken into account and corrected using photometric data.  
In section 4,  we show the resulting U-V distribution for the whole dataset and for subsamples inside and outside the solar circle ($R_{\rm 0}\simeq 8$ kpc). In section 5, we discuss the new features revealed in the U-V distribution and give possible explanations for their existence. We conclude our investigation in Section 6.

\section{Extreme deconvolution}\label{sect:deconv}

We derive the intrinsic distribution of the LAMOST stars in the U-V plane using the extreme deconvolution method described in \citet{Bovy09}.  In this section, we summarise the key points of the deconvolution method relevant to our investigation.

A multiple Gaussian model is applied as the empirical model of the U-V distribution of a group of stars:
\begin{equation}
p(\vv)=\sum_{j=1}^{\rm K}{{A_j}\mathcal{N}(\vv|\muj,\Sigmaj)},
\end{equation}
where K is the number of Gaussians, the amplitudes $A_j$ sum to unity and $\mathcal{N}(\vv|\muj,\Sigmaj)$ the $j$th multivariate Gaussian distribution with mean $\muj$ and covariance matrix $\Sigmaj$.
 Given a star $i$, the observed velocity, $\hatvi$, is the sum of the true velocity, $\vvi$, and a random error, $\ei$ induced during the measurement:

\begin{equation}
\hatvi=\vvi+\ei.
\end{equation} 
In general, $\ei$ follows a Gaussian distribution with zero mean and covariance matrix of $\Si$. The probability of the observed velocity, $\hatvi$, given the model parameter, $\theta$, can be derived from the true distribution of velocity, $p(\vv)$, convolved with the measurement error:
\begin{equation}
p(\hatvi|\theta)=\sum_{j=1}^{\rm K}{{A_j}\mathcal{N}(\hatvi|\muj,\Tij}),
\end{equation}
where 
\begin{equation}
\Tij=\Sigmaj+\Si.
\end{equation}
The true distribution model can be numerically solved using an extreme deconvolution algorithm, which removes the effect of the uncertainty of the velocity estimates \citep{Bovy09}.

For a fixed number of Gaussians, K, the likelihood of the multi-Gaussian model given a set of parameters can be expressed as
\begin{eqnarray} \label{eq:likelihood}
\mathcal{L}= \sum_{i} W_i \, {\ln}\ p(\hatvi|\theta) = \sum_{i} W_i \ln\sum_{j=1}^{\rm K}A_j\mathcal{N}(\hatvi|\muj,\Tij) .
\end{eqnarray}
where $W_i$ is the weight of the i-th star.
An Expectation-Maximization (EM) technique is then used to find the parameters maximizing the likelihood \citep{Bovy09}.

\begin{figure*}
	\centering
 	\includegraphics[width=0.9\textwidth]{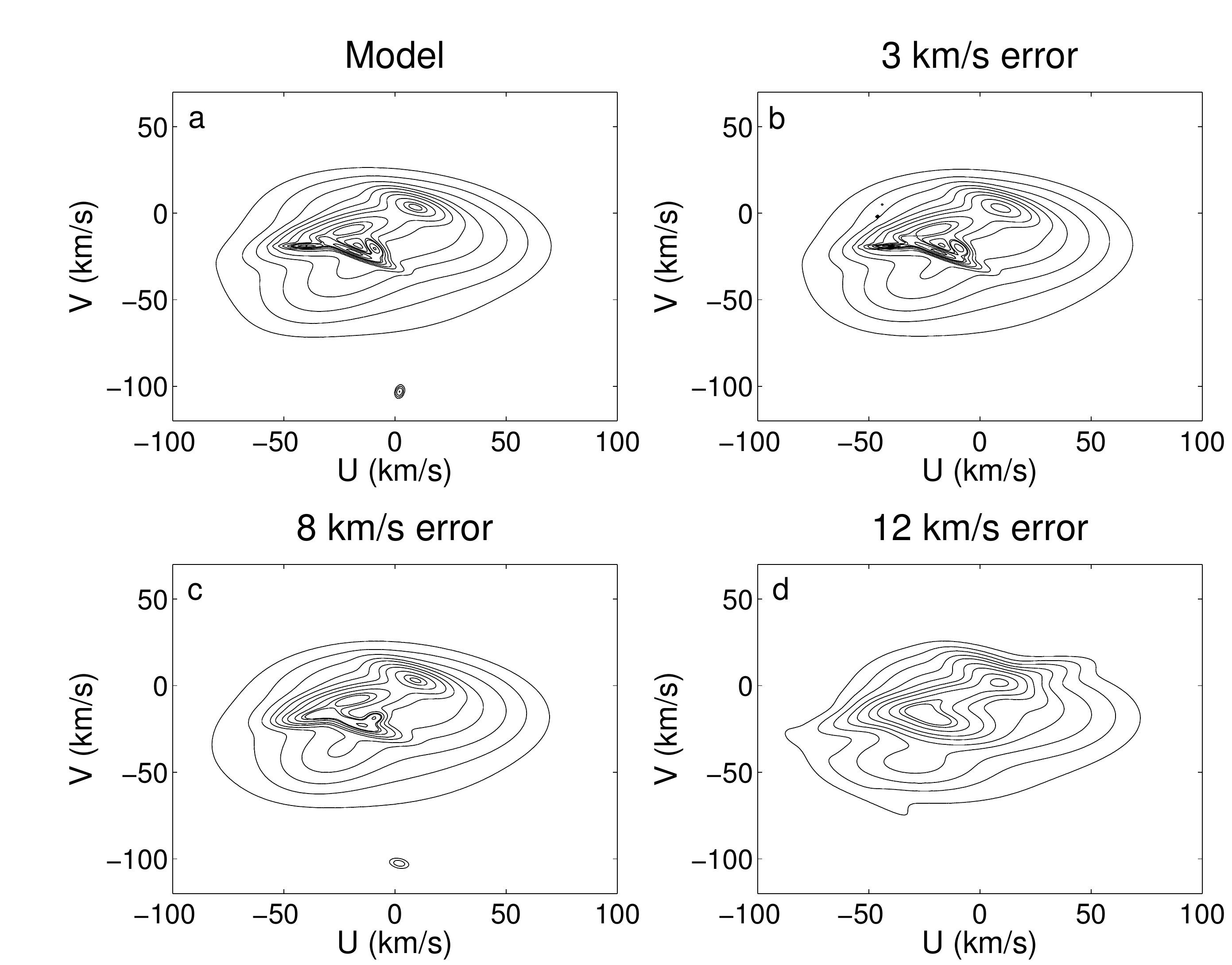} 
 	\caption{The mock U-V velocity distribution. The contours contain, from inside outward, 1\%, 3\%, 8\%, 13\%, 25\%, 40\%, 48\%, 65\%, 75\%, 83\%, 89\% and 94\% of stars. }
	\label{fig:sim}
\end{figure*}

In principle, the velocity error may affect the performance of the extreme deconvolution. Small and fine structures in the velocity distribution may not be recovered by the extreme deconvolution if the error of the observed velocity is too large. In theory, structures with scale smaller than the error are unreliable. So we set the regularization parameter $w$ \citep{Bovy09} roughly the same as the square of the error. We perform Monte Carlo simulations to investigate this effect.

We select 10 Gaussians as the original distribution using the parameters listed in Table 1 of \citet{Bovy09}. Because only U and V are considered in our work, the dimension W (vertical velocity) of these Gaussians is ignored. The true U-V distribution of the 10-Gaussian model is shown in Figure~\ref{fig:sim}a. The scales \footnote{In this paper, the scale of a 2-D Gaussian is defined as the square root of the trace of the covariance matrix.} of the components vary from 2.3\,\kms\ to 95\,\kms. Notice that the scales of the three compact Gaussian components located between $\rm U\sim-50$ and $0$\,\kms\ at $\rm V\sim-20$\,\kms\ are 5.6\,\kms, 9.3\,\kms, and 4.7\,\kms, respectively, from the left to right. The scales of the Gaussians at (U, V)$\sim(-22, -10)$\,\kms\ and (9,  4)\,\kms\ are 17\,\kms\ and 10\,\kms, respectively.
To create the data, we randomly generate 20,002 mock stars from the original 10-Gaussian distribution, and with additional arbitrary Gaussian errors as the test data. The U-V distribution is reconstructed using a 12-Gaussian empirical model with the extreme deconvolution method. Figures~\ref{fig:sim}b, c, and d show the results with the Gaussian errors at 3\,\kms, 8\,\kms\ and 12\,\kms, respectively. The corresponding regularization parameters $w$ are chosen to be 9\,\kmss, 64\,\kmss, and 144\,\kmss. As mentioned before, we set $w$  comparable with the velocity uncertainties of the data. If $w$ is set with a value smaller than the velocity uncertainty, smaller structures can be revealed, however, they are usually spurious. On the other hand, larger structures should not be affected by the smoothing, provided that the value of $w$ is not too large. Therefore, we select $w$ to be comparable to the velocity uncertainty squared.

For the test with a random error of 3\,\kms, Figure~\ref{fig:sim}b shows that almost all of the substructures in the original distribution are reconstructed, except the low-amplitude one at $(0, -100)$\,\kms. The random initial conditions of the extreme deconvolution and relatively fewer Gaussian components may lead to the loss of small substructures.
For the test with random error of 8\,\kms, Figure~\ref{fig:sim}c shows that the three most compact structures around V$\sim-20$ are not distinguishable and turn into a single larger-scale substructure. When the measurement error goes up to 12\,\kms, the component at $(-22, -10)$\,\kms, together with the three most compact substructures, is merged into a single larger substructure centred around $(-25, -15)$\,\kms. The larger component at (9, 4)\,\kms\ is still distinguishable, but an elongated Gaussian component is added to connect it with other substructures, making an artificial ridge from $(-80, -30)$ to $(50, 15)$\,\kms.
For the test case of the 10-Gaussian U-V distribution, which mimics the true distribution in the solar neighbourhood, when the measurement error of the velocity is as high as 12\,\kms, the extreme deconvolution can not properly recover smaller-scale substructures while some artifacts may also affect the larger scale components in terms of their shapes.

\section{Observational Data}\label{sect:data}

LAMOST (also called Guoshoujing telescope) is a 4 meter quasi-meridian reflective Schmidt telescope aiming for about 5 million spectra in a 5-year survey \citep{Cui12}. The LAMOST survey will provide important stellar spectroscopic observations for a broad range of studies on the Milky Way \citep{Deng12}. Stars located in the solar neighbourhood are the perfect sample for studying the stellar velocity distribution. The LAMOST pilot survey has observed about 640,000 stellar spectra from October 2011 to June 2012 \citep{Zhao12}. Because the LAMOST stellar parametrization method is still in development, only the radial velocities of the stars have been released. As a consequence, we use the 2MASS magnitudes \citep{Cutri03} of the LAMOST targets to select the F/G dwarfs stars for this work and estimate their distances based on their color indices. We cross-identify the LAMOST stellar spectra with 2MASS photometry using their positions. A small fraction of faint stars in the LAMOST dataset may be beyond the limiting magnitude of 2MASS and thus are missed during the cross identification. However, since we focus only on the solar neighbourhood, these missing faint stars should not affect our results. The interstellar extinction in the $J$, $H$, and $K$ bands is estimated using the $E(B-V)$ map from \citet{Schlegel98}. We adopt the extinction coefficients in J, H, K and V as
 \begin{equation}
 \begin{array}{cc}
 A_J=&0.27A_V,\\
 A_H=&0.17A_V,\\
 A_K=&0.11A_V,
 \end{array}
 \end{equation} 
following \citet{Fiorucci03} and \citet{Girardi04} with $R_V=3.1$. The extinction $A_J$, $A_H$, and $A_K$ is calculated for each star and removed from the apparent magnitudes. The F and G stars are selected with $0.2<(J-K)_0<0.45$, where $(J-K)_0$ is the dereddened color. In total 39,862 stars with $|b|>30^\circ$ are selected for this work. Because F and G giant stars are very rare, we ignore them and treat the whole sample as main-sequence stars. 

The proper motions of the sample are obtained by cross-identifying with the PPMXL catalog \citep{Roser10}. It is noted that the systematic bias of the proper motions in PPMXL is not negligible \citep{Wu11}. Therefore, we use QSOs (Quasi-stellar object) to correct the bias on a star-by-star basis  (see Appendix A for details).
 
\subsection{Distance estimation}\label{sect:distance}
\begin{figure}
\begin{center}
\includegraphics[scale=0.45]{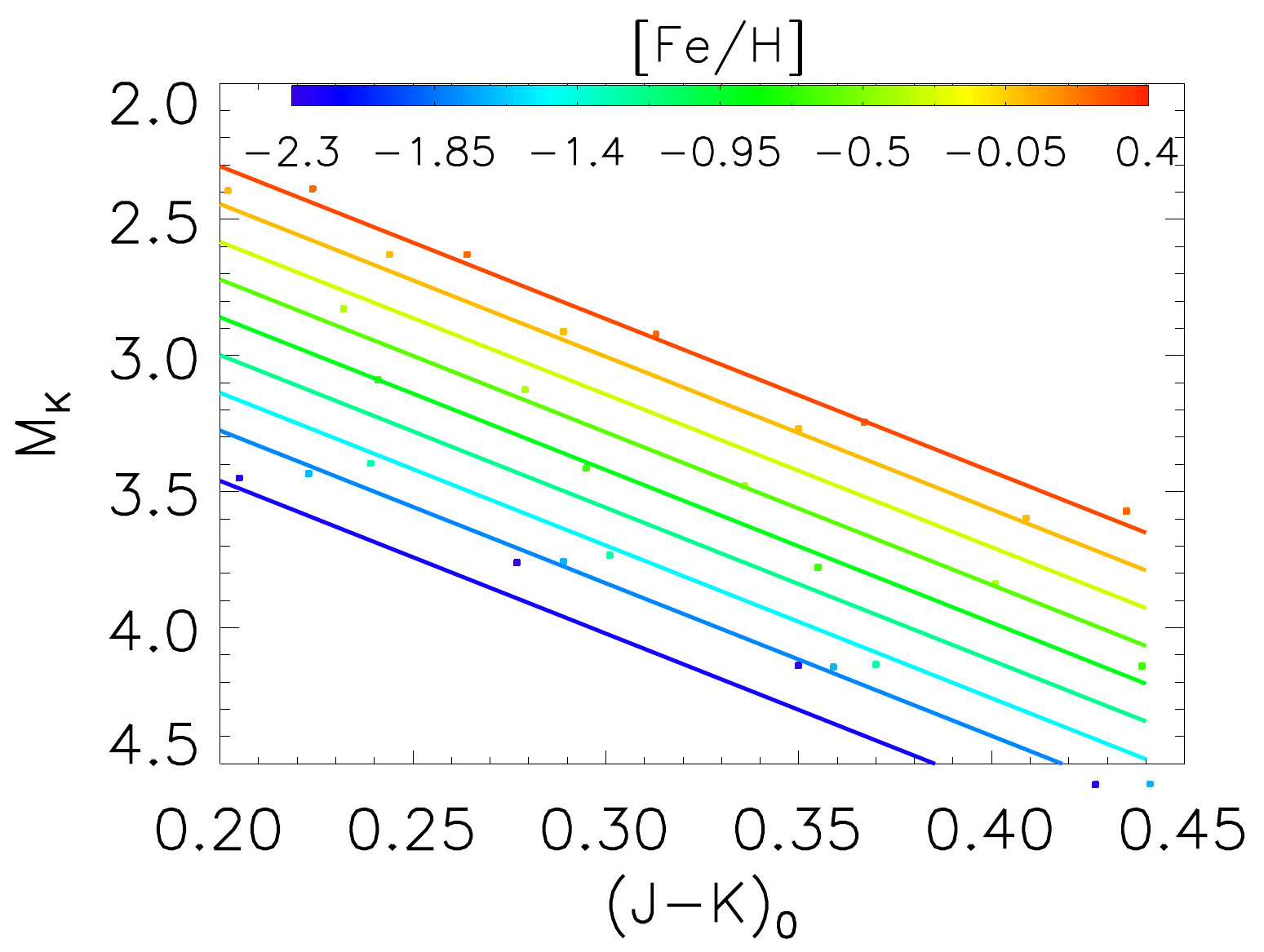}
\caption{The absolute magnitude in the $K$ band, $M_K$, as a function of $(J-K)_0$ and [Fe/H]. The dots show the synthetic samples from isochrones with $\log({\rm age/yr})=7.85$. The color encodes the metallicity. The series of lines show the best surface plane fit for this relation.}\label{fig:mkjkfeh}
\end{center}
\end{figure}

\begin{figure}
\begin{center}
\begin{minipage}{8.5cm}
\includegraphics[scale=0.45]{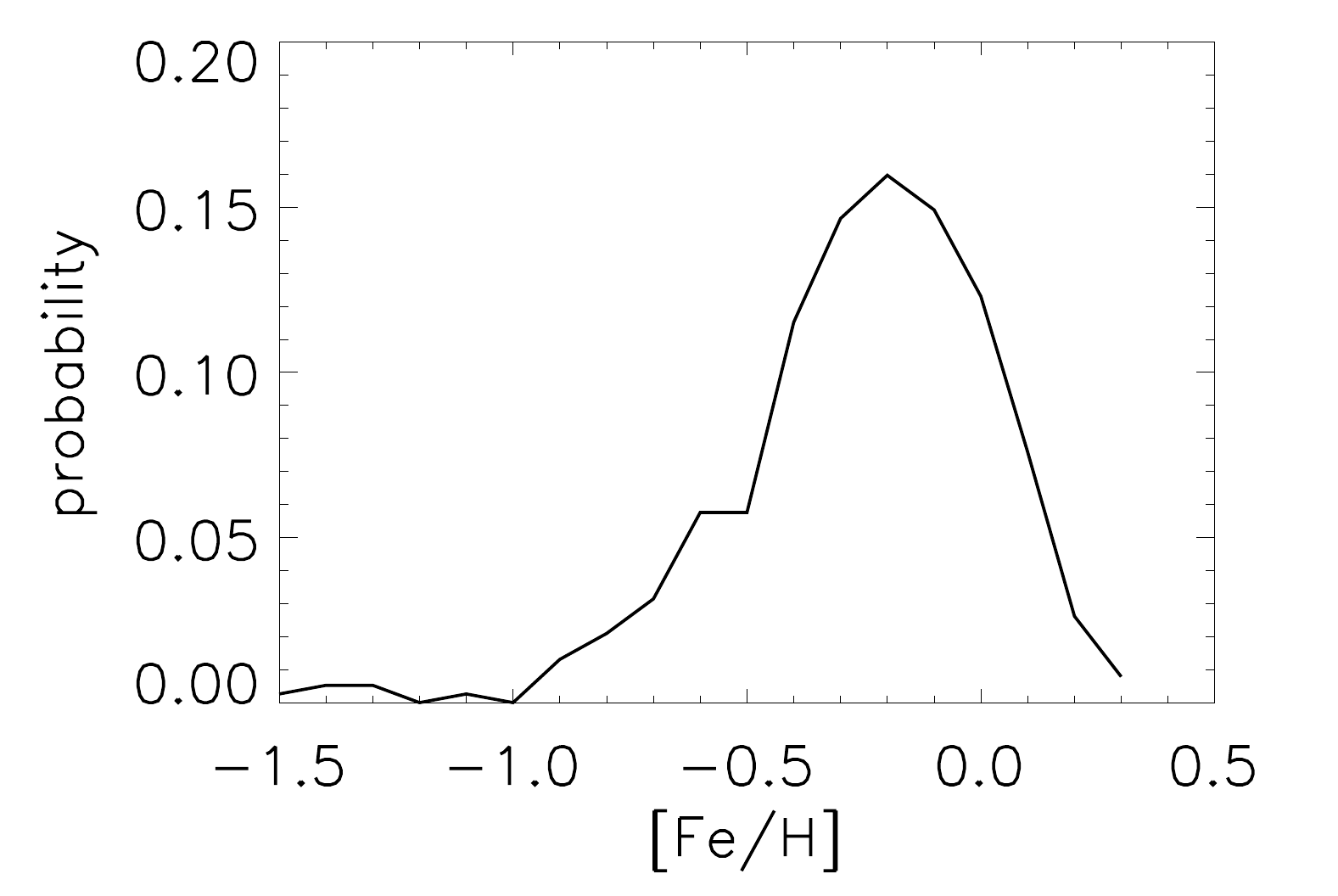}
\includegraphics[scale=0.45]{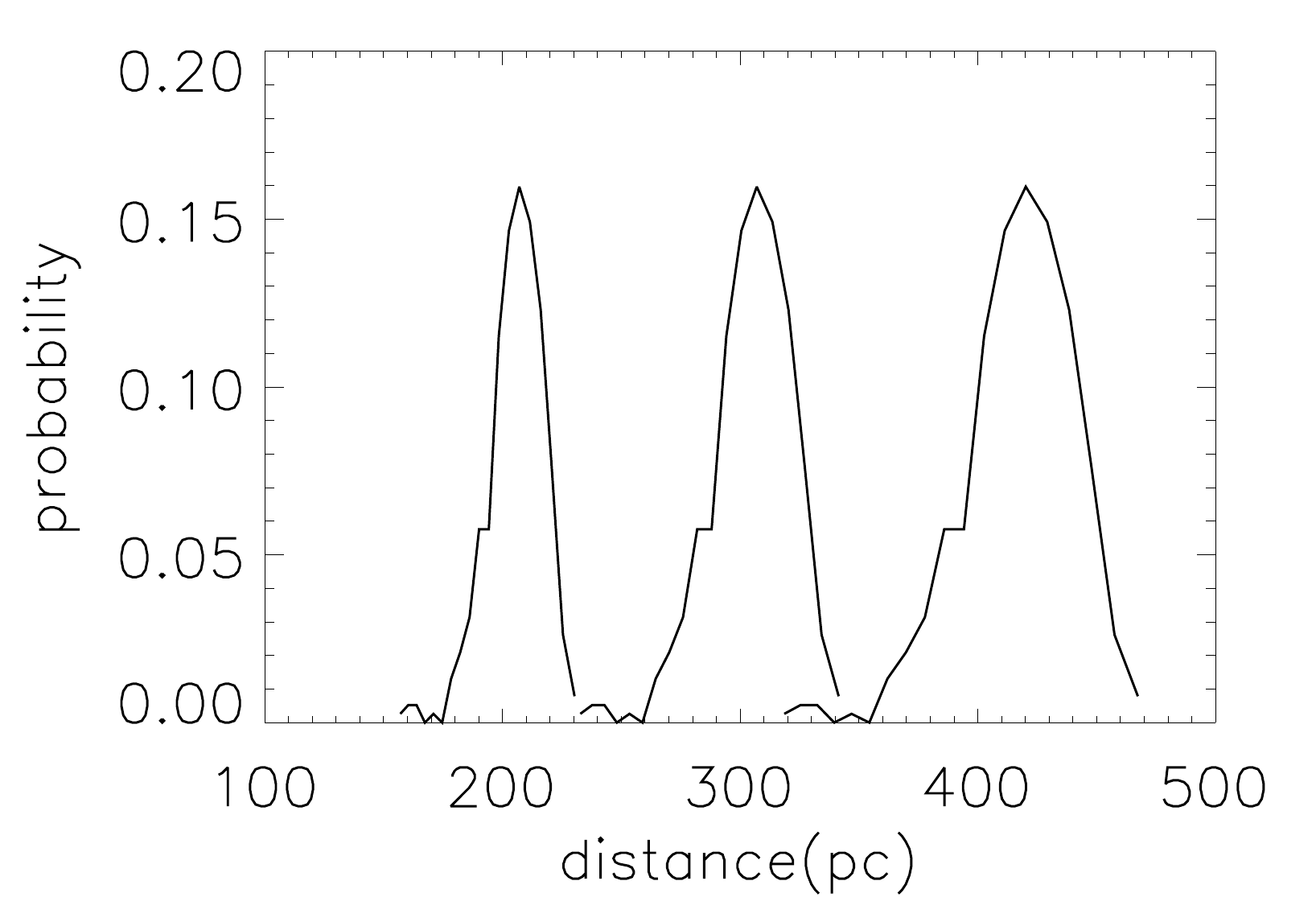}
\end{minipage}
\caption{\textit{Top panel: } The metallicity distribution function (MDF) for G dwarf stars in the solar neighbourhood \citep{Hou98}. \textit{Bottom panel:} The probability density functions (PDFs) of distance for three samples of F/G dwarf stars at various distances. The PDFs are derived from the MDF shown in the top panel.
}\label{fig:distancepdf}
\end{center}
\end{figure}

\begin{figure}
\centering
 \includegraphics[width=3in]{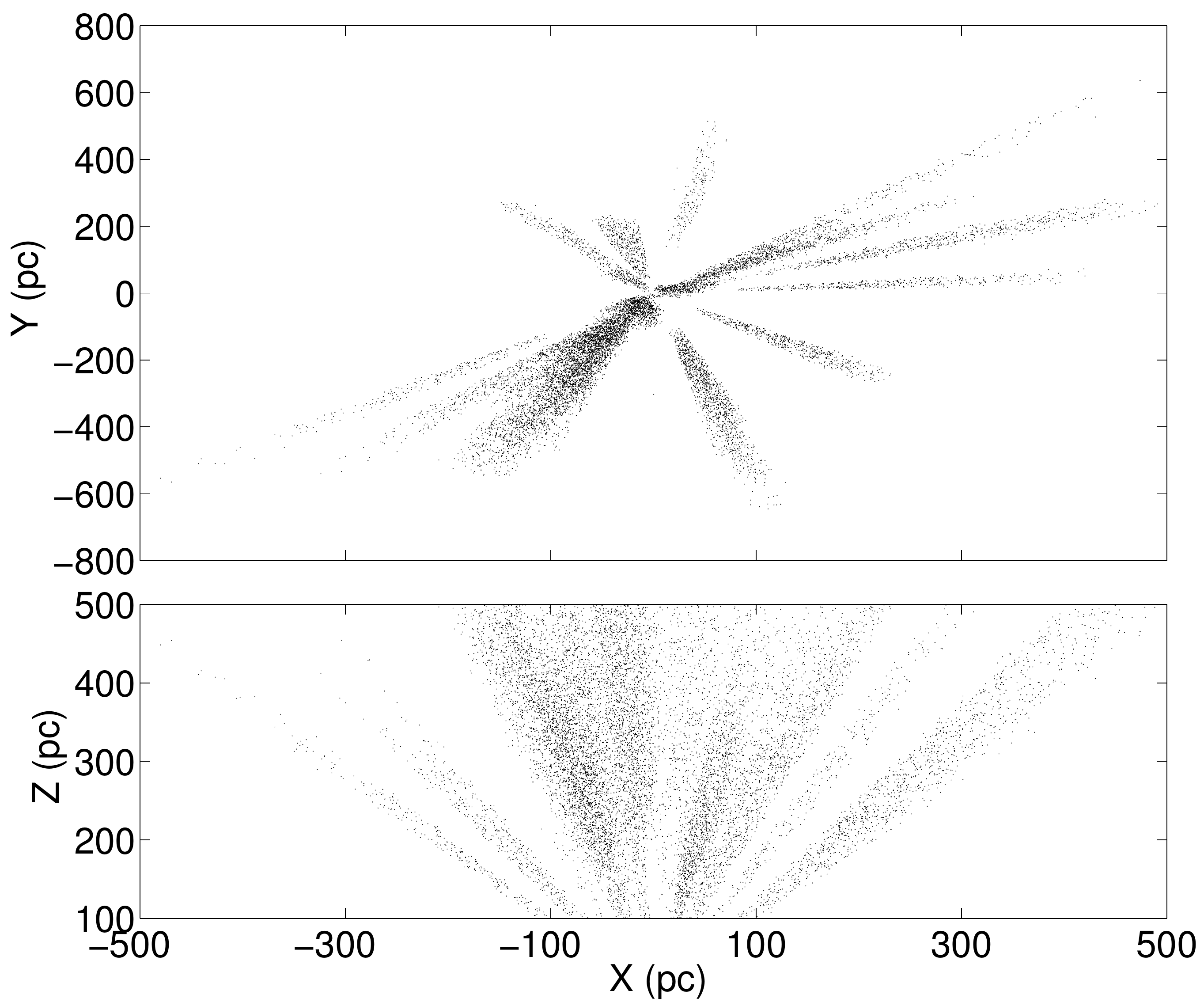} 
 \caption{The locations, in the galactic coordinate system, of the stars we used. The direction of the positive $x$-axis is toward the Galactic centre.}\label{fig:location}
\end{figure}

\begin{figure}
\centering
\begin{minipage}{8.5cm}
\includegraphics[width=3in]{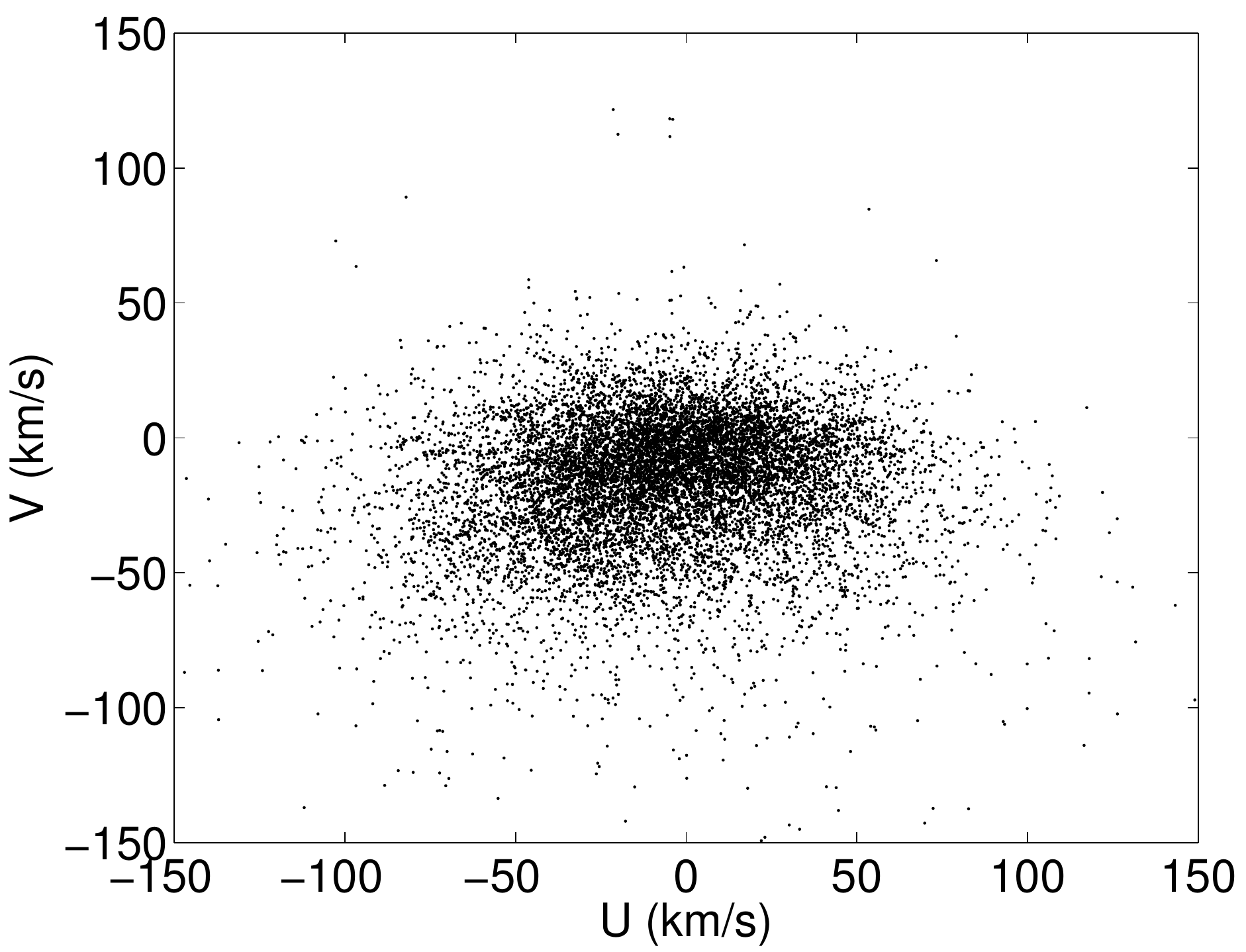} 
\includegraphics[width=3in]{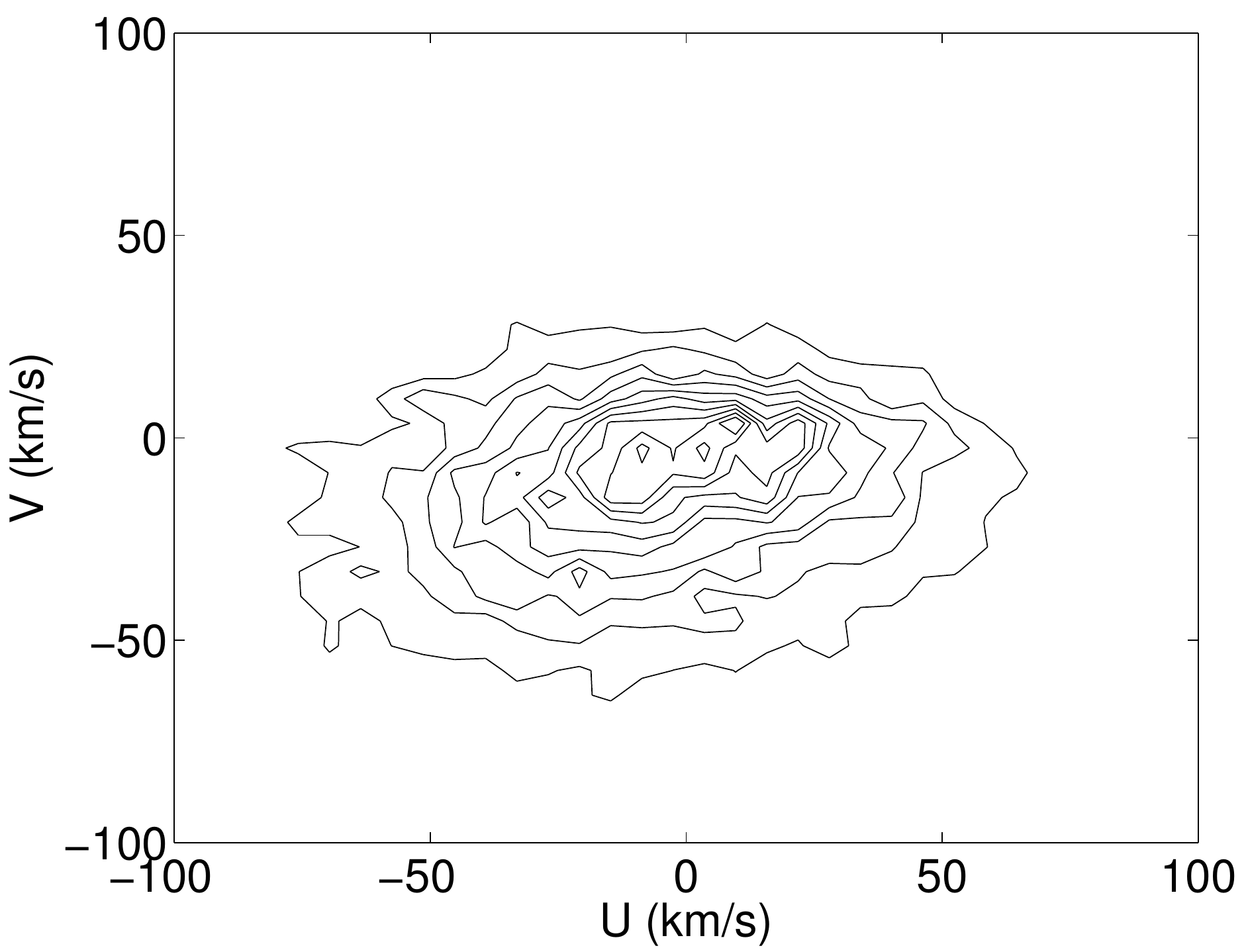} 
\end{minipage}
\caption{\textit{top panel:} Scatter plot of the velocity distribution in the U-V plane. \textit{bottom panel:} The contour of the velocity distribution in the U-V plane. The contours contain, from inside outward, 3\%, 8\%, 19\%, 30\%, 39\%, 50\%, 59\%, 71\%, 85\% of stars.}\label{fig:meanU-Vdist}
\end{figure}

The distance of the F/G dwarf stars in the sample can be estimated from the color index and metallicity. Figure~\ref{fig:mkjkfeh} shows the absolute magnitude, $M_K$, as a function of $(J-K)_0$ and [Fe/H] in the isochrones given that $\log({\rm age/yr})=7.85$ \citep{Girardi02}.
We use the relatively young isochrones here because they cover the full range of the color index $(J-K)_0$. However, this may induce a systematic bias in the distance estimation. In principle, the young population should be a little fainter than an old one on the main-sequence. This may lead to an underestimation of the distance by a factor of 20\%, which would produce an under-estimate of U and V by a similar percentage. Since these systematics rescale U and V by similar factors for most of the stars, they will not significantly change the relative distribution of the stars in the U-V plane.

Therefore, we select the oldest age with main sequences that can cover the full range of $0.2<(J-K)_0<0.45$.
We fit the $M_K((J-K)_0,\rm{Fe/H])}$ with a 2-D plane:

\begin{equation}
M_K=a_0+a_1(J-K)_0+a_2\rm{[Fe/H]},
\end{equation}
and find the best-fit coefficients are $a_0=1.32194$, $a_1=5.6081$, and $a_2=-0.462324$.

The LAMOST pilot survey does not provide [Fe/H], and hence we can only obtain the probability density function (PDF) of $M_K$ for a star given a metallicity distribution function (MDF) in the solar neighbourhood. \citet{Hou98} (see also the top panel of Figure~\ref{fig:distancepdf}) provide an MDF for G dwarf stars in the solar neighbourhood, which is suitable for our sample. Although the MDF may shift to the metal-poor end with increasing height ($z$) over the Galactic mid-plane, it can only introduce at most a 4\% difference in the distance estimation at $z=0.5$\,kpc given that the vertical metallicity gradient is $-$0.3\,dex kpc$^{-1}$. This subsequently leads to at most a 2\,\kms\ shift in the tangential velocity (V) when the true velocity is 50\,\kms. Therefore, we simply assume the MDF is constant with $z$ in this paper. The PDFs of distance for three sample stars are shown in the right panel of Figure~\ref{fig:distancepdf}.

Figure~\ref{fig:location} shows the 3-D positions in heliocentric Cartesian coordinates for the 14,662 selected F/G dwarf stars based on their mean distance estimates. We select stars between 100 and 500\,pc in $z$ for two reasons. First, stars located nearer than 100\,pc are not completely sampled in the LAMOST pilot survey since they are too bright. This may lead to some strong selection effects. Second, the stars located farther away than 500\,pc are dominated by luminous stars and thus lead to selection effects in the opposite way. Therefore, we select stars within this range to keep the luminosity function approximately constant at different $z$.

The velocity components U and V in heliocentric Cartesian coordinates are calculated according to \citet{Johnson87}. Because the distance of a star in this work follows a certain PDF, the U and V are also  random variables based on the distance PDF. Figure~\ref{fig:meanU-Vdist} shows the distribution of the mean U and V derived from the mean distances. The Hercules stream, which leads to an obvious asymmetry in U is clearly displayed in the figure. More detailed substructures will be unveiled in section~\ref{sect:results} after applying the extreme deconvolution algorithm to the data.

\subsection{Selection correction}\label{sect:selection}
\begin{figure}
\centering
 \includegraphics[width=3in]{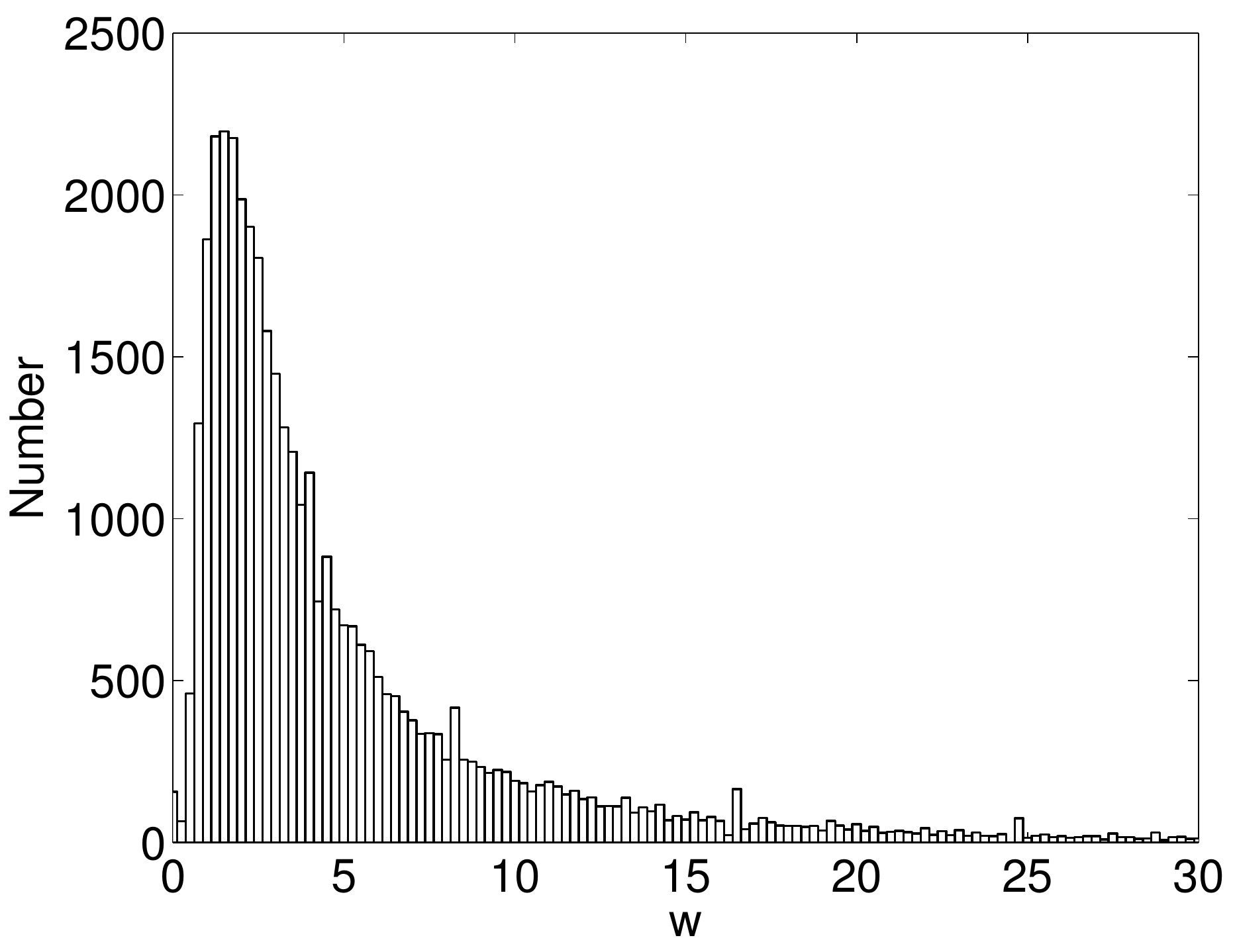} 
 \caption{The histogram of the selection correction weights (see the text in Section \ref{sect:selection}). Particles with $0.5\le \varpi \le 6$ are used in our study.}\label{fig:weight}
\end{figure}

In general, the sampling of a spectroscopic survey is significantly affected by the selection strategy, observational conditions, data reduction, etc. These may induce selection bias. In order to reduce the selection effects in the resulting U-V distribution, we use photometric data to correct the bias.

The 2MASS photometric survey is assumed to be a complete dataset for F/G dwarf stars given a range of magnitudes. The stellar count within a small solid angle, $d\Omega$, a small range of $K$ magnitude, $dK$, and a small range of color index, $d(J-K)_0$, around a given star can be written as 
\begin{equation}
\begin{aligned}
N_{\rm{ph}}&(\alpha,\delta,K,(J-K)_0)=\\
&\nu(\alpha,\delta,K, (J-K)_0)\,d\Omega \, dK\, d(J-K)_0,
\end{aligned}
\end{equation}
where $\alpha$ and $\delta$ are the central right ascension and declination of $d\Omega$, and $\nu$ is the stellar density in the small volume. For the LAMOST spectroscopic data, we have
\begin{equation}
\begin{aligned}
N_{\rm{sp}}&(\alpha,\delta,K,(J-K)_0)=\\
&{1\over{\varpi}}\nu(\alpha,\delta,K, (J-K)_0)\,d\Omega \,dK\,d(J-K)_0,     
\end{aligned}                                                                                 
\end{equation}
where $1/\varpi$ is the selection function of the survey and $\varpi$ is the weight of the star. For our data, $dK$ is set to 1 magnitude, $(J - K)_0$ is 0.2 magnitude and the solid angle $d\Omega$ is 0.5 sr. The weight $\varpi$ can be derived from the ratio of the photometric stellar count to that of the spectroscopic survey. When $\varpi \sim1$, the spectroscopic survey essentially samples all the stars with similar direction, magnitude, and color. When $\varpi \gg1$, the spectroscopic data is under-sampled. In general, $\varpi$ should not be less than 1 unless some stars have more than one spectroscopic observation. This weight can be used to correct the stellar count for the spectroscopic survey data in equation~\ref{eq:likelihood}. 

Figure~\ref{fig:weight} shows the distribution of $\varpi$ for the dataset. The peak value of $\varpi$ is around 2, meaning that each star in the spectroscopic samples represents usually two photometric ones with similar position, magnitude, and color index. The stars with $\varpi$ larger than 6 (about $20\%$ of the total number of stars) have been removed from our sample because they are highly under-sampled and may not therefore represent the kinematic features for similar stars. Furthermore, stars with $0.5\le \varpi<1$ are also excluded. They may either be observed more than once, or have a smaller value of $N_{\rm sp}$ due to very few stars counted in the given volume and hence are not suitable for  later statistical studies. 

\section{Results}\label{sect:results}

\begin{figure}
\centering
 \includegraphics[scale=0.35]{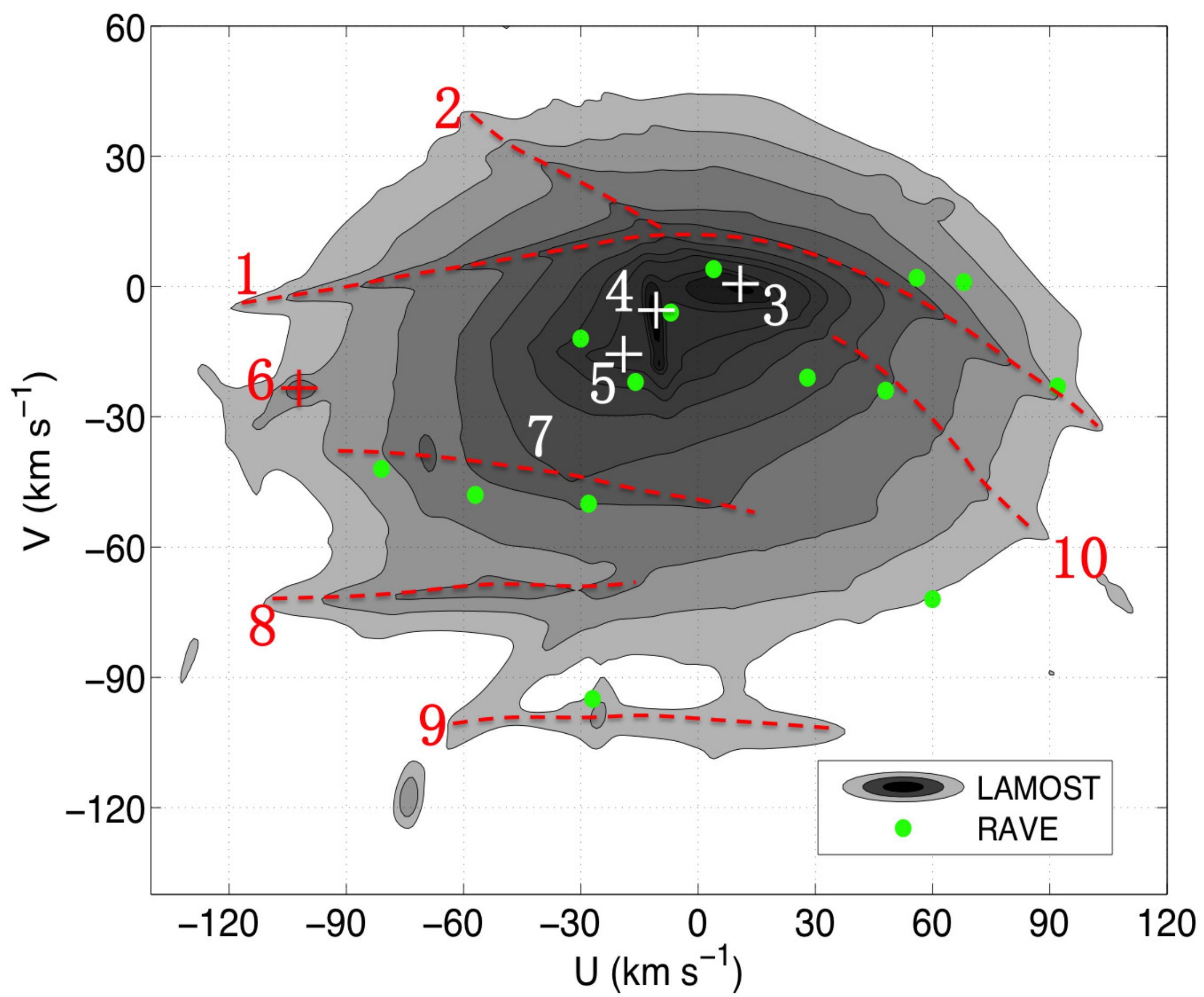} 
 \caption{The figure shows the velocity distribution in the U-V plane. The contours contain, from inside outward, 2\%, 6\%, 12\%, 21\%, 32\%, 48\%, 58\%, 70\%, 81\%, 88\%,  94\% and 97\% of stars. The circles are the central positions of  the over-densities in RAVE \citep{Antoja12}. The identified structures are marked as crosses or dashed lines with a number aside. The detailed locations and the names of the known substructures are listed in Table~\ref{tab:substructures}.}\label{fig:uv}
\end{figure}

\begin{figure}
\centering
\begin{minipage}{8.5cm}\begin{center}
 \includegraphics[scale=0.35]{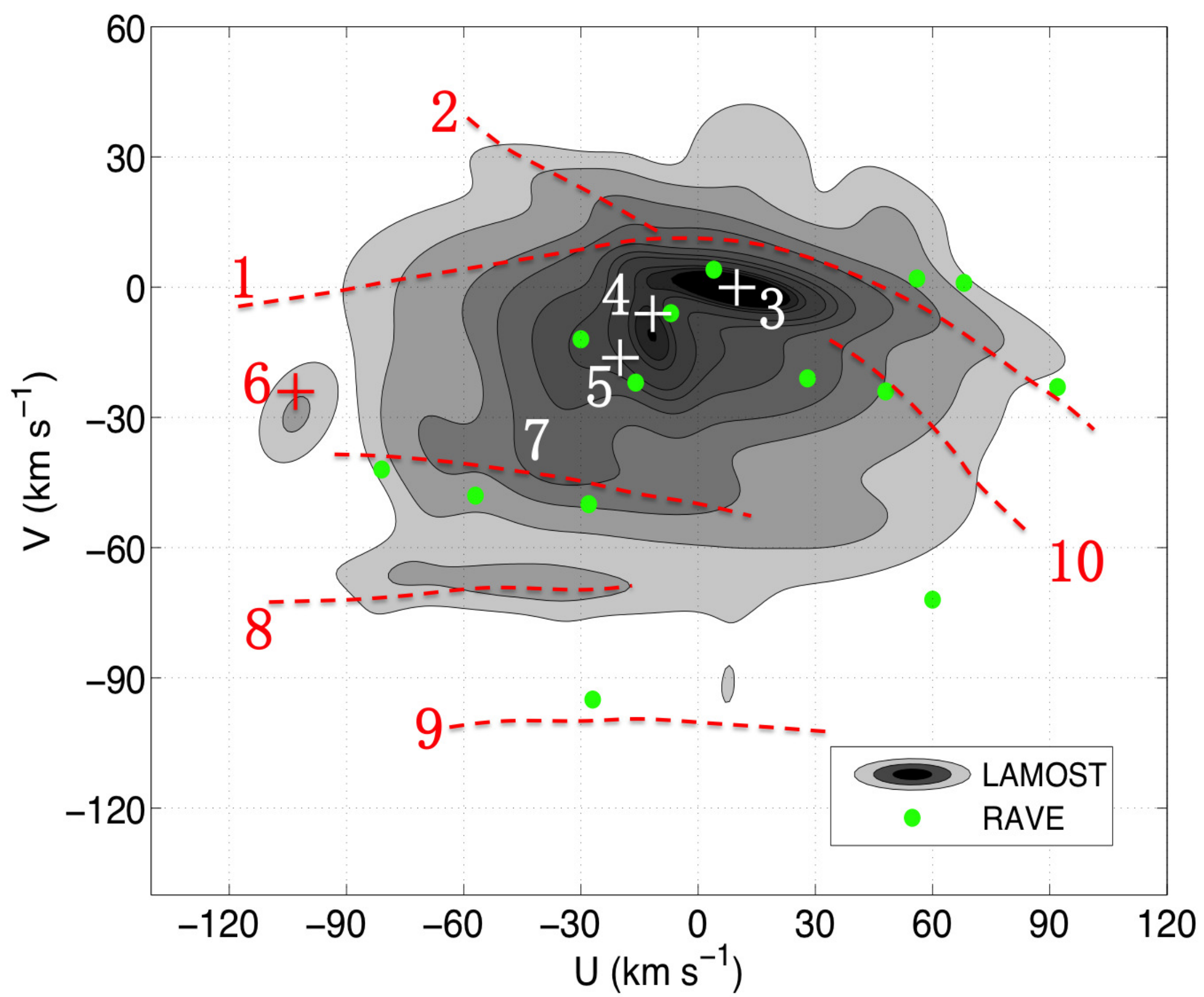}
 \includegraphics[scale=0.35]{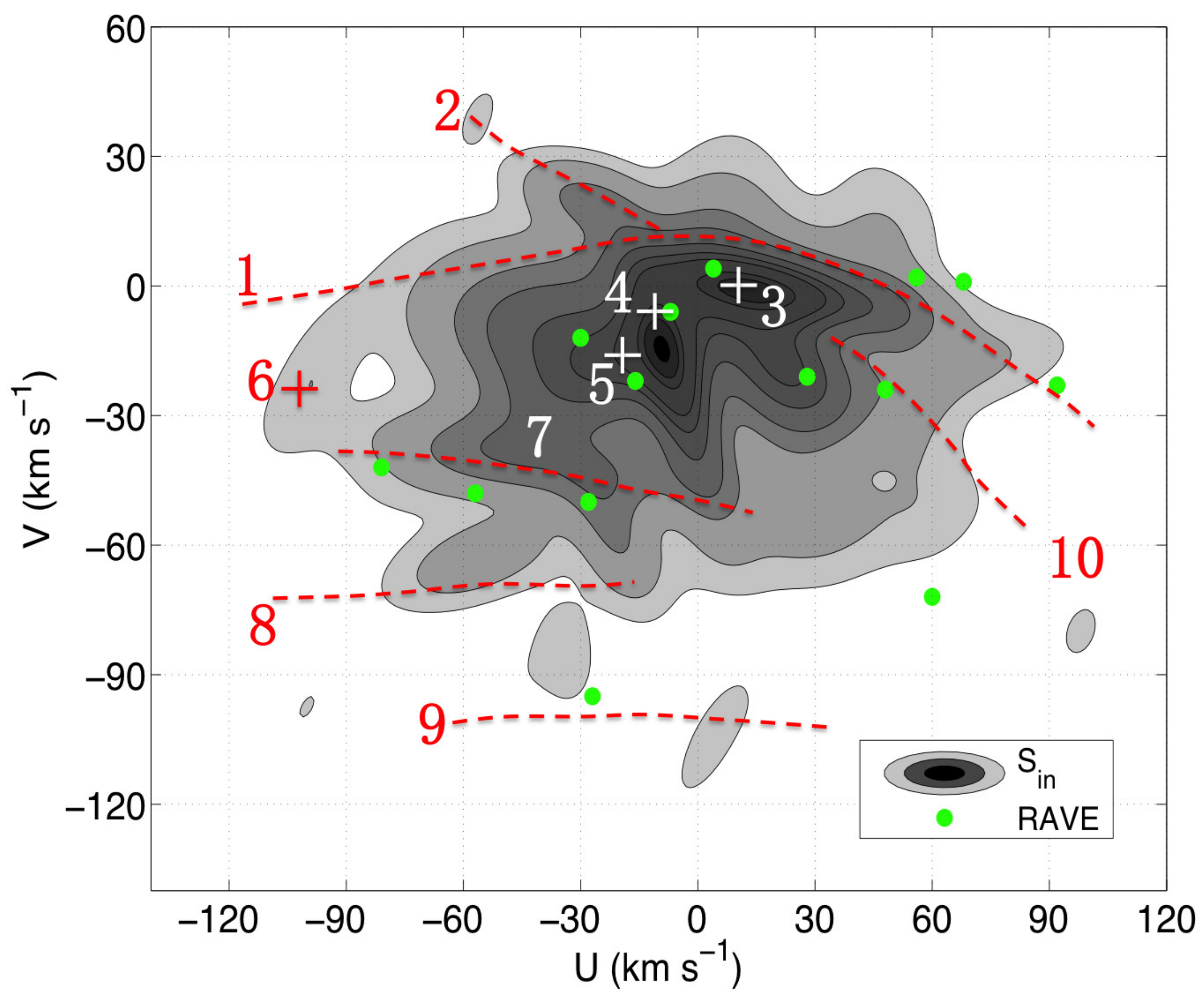}
 \includegraphics[scale=0.35]{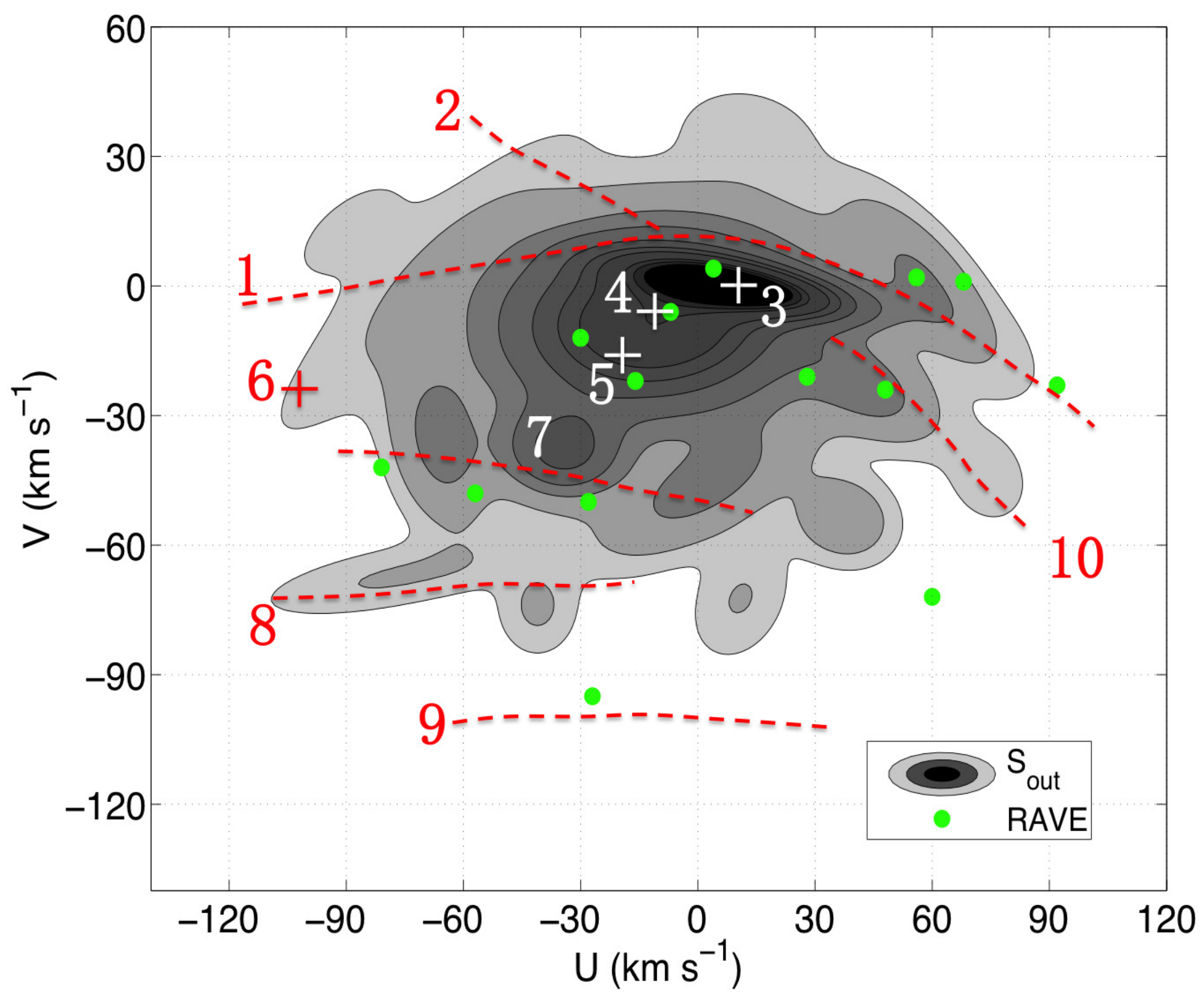} 
 \caption{The top panel shows the U-V distribution using the mean distances. The middle panel is the velocity distribution of the $S_{\rm in}$ sample (for stars within the solar circle, $R<R_{\rm 0}$) and the bottom panel is the velocity distribution of the $S_{\rm out}$ sample (for stars beyond the solar circle, $R>R_{\rm 0}$). The contour levels are the same as in Figure~\ref{fig:uv}.}\label{fig:meanuv}
\end{center}
\end{minipage}
\end{figure}

\begin{table}
\centering
\caption{The list of substructures unveiled from Figure~\ref{fig:uv}}\label{tab:substructures}
\begin{tabular}{clcc}
\hline\hline
\small{No.} & \small{Name} & \small{U range} & \small{V range}\\
\hline
\small{1} & \small{NEW} & \small{$-120\sim103$} & \small{$-5\sim-32$} \\
\small{2} & \small{NEW} & \small{$-60\sim-15$} & \small{$40\sim15$} \\
\small{3} & \small{Sirius} & \small{11} & \small{-1} \\
\small{4} & \small{Coma Berenices} & \small{-11} & \small{-7} \\
\small{5} & \small{Hyades-Pleiades} & \small{-18} & \small{-18} \\
\small{6} & \small{NEW} & \small{-102} & \small{-24} \\
\small{7} & \small{Hercules} & \small{$-95\sim5$} & \small{$-38\sim-50$} \\
\small{8} & & \small{$-111\sim-14$} & \small{$-73\sim-68$} \\
\small{9} & \small{Arcturus} & \small{$-64\sim37$} & \small{$-100\sim-102$} \\
\small{10} & \small{Wolf 630} & \small{$33\sim90$} & \small{$-11\sim-58$} \\
\hline
\end{tabular}
\end{table}

\subsection{Reconstruction of the intrinsic U-V distribution}\label{sect:reconst}
In order to reconstruct the U-V distribution for the selected samples, Monte Carlo simulations combined with the extreme deconvolution are used. First, a random distance for each star is drawn from the distance PDF and then a pair of corresponding U and V is determined. Second, an extreme deconvolution with 20 Gaussians is applied to such a dataset in the U-V plane. Comparing the model predicted radial velocities to those of the stars from the GCS catalog, \citet{Bovy09} inferred that 10 Gaussians can work well to reconstruct the U-V-W distribution with \textit{Hipparcos} data. For our case, we use 20 Gaussians in our model in each random draw according to Appendix B. Since the velocity error is around 6 \kms, the regularization parameter $w$ in this fitting is chosen to be 36 \kmss.

We run 100 random draws and derive the median U-V distribution over the 100 20-Gaussian models. We compare the median U-V distribution to that averaged over the first 50 draws and find that the main structures do not significantly change. Therefore, we believe that the results from 100 draws have converged and hence are sufficient for the reconstruction of the U-V distribution. 

We also run the extreme deconvolution with average distance to estimate the U-V distribution in order to save computational time (see Figure~\ref{fig:meanuv}). However, We argue below that the Monte Carlo simulations used here can lead to better resolution in the U-V distribution and therefore is able to reveal more detailed features than simply using the average distance.

First, the Monte Carlo simulations can take into account the non-Gaussian profile of the uncertainty in distance, while the extreme deconvolution with the average distance can only approximate it as a Gaussian and may miss some information due to the non-Gaussian profile. Second, the dispersion in U-V distribution due to the uncertainty in distance in the Monte Carlo simulations should be equivalent with that in the extreme deconvolution with average distance. However, because the uncertainty of the average distance is propagated to U and V, their uncertainties should be slightly larger than a single realisation in the Monte Carlo process. The 10\% distance uncertainties generate additional 5 \kms\ velocity errors which increase the velocity errors from about 6 \kms\ to 8 \kms\ in both U and V. Therefore we use a slightly larger regularization parameter, $w=64$ (km/s)$^2$, in the extreme deconvolution with average distance than that in the Monte Carlo processes, for which we take $w=36$ (km/s)$^2$. The larger regularization may lead to a slightly smoother distribution but smaller substructures will be blurred. Finally, in general, the true U-V distribution may contain more substructures than the Gaussian components of an empirical model. It implies that the extreme deconvolution with a limited number of Gaussians actually smooths out the U-V distribution and may miss some subtle substructures due to the lack of Gaussian components to represent for them. However, in the Monte Carlo simulation, for each realization, we first apply 20 Gaussians in the empirical U-V distribution model and finally apply 100$\times$20 after 100 simulations. Recall that the regularization parameter is smaller in the Monte Carlo simulations, which implies that it is more sensitive to smaller substructures than the extreme deconvolution method. Therefore, the Monte Carlo simulations not only have far more Gaussian components to represent the substructures, they also have an advantage in revealing smaller substructures. Although averaging the 100$\times$20 Gaussians may eventually smooth out some substructures, nevertheless, the reproductions of small substructures in some realizations will still remain in the final distribution. We conclude that the Monte Carlo simulations (shown in Figure~\ref{fig:uv}) can reveal more details than the average distance approach in Figure~\ref{fig:meanuv}.

\subsection{Known over-densities in the LAMOST data}
Figure~\ref{fig:uv} shows our results. Table~\ref{tab:substructures} lists all the identified substructures in Figure~\ref{fig:uv}. Unlike the results based on the wavelet transform (e.g., \citealt{Chereul99, Zhao09, Antoja12} etc.), the LAMOST pilot survey data shows very smooth structures in the U-V distribution, except for the two compact over-densities in the core of the distribution, i.e., structures 3 and 4, which are the Sirius and Coma Berenices structures, respectively.  The centre of the two over-densities is a few \kms\, offset from \citet{Antoja12} (shown as the circles in the figure). This is probably because the velocities between the RAVE and LAMOST data are not well calibrated, especially the contribution from the proper motions.

Hyades and Pleiades over-densities are connected with each other and marked as structure 5 in Figure~\ref{fig:uv}. The two over-densities may be parts of one single elongated substructure at $\rm V\sim-20$\,\kms. Indeed, \citet{Famaey05} did not separate them using CORAVEL/\textit{Hipparcos}/Tycho-2 data. They can be separated in studies based on the wavelet transform since it tends to remove the low-frequency (or smooth) components and enhance the high-frequency (or clumpy) components. On the other hand, it is unlikely that they are not distinguished by our extreme deconvolution method because of the uncertainty of the velocity. \citet{Dehnen98} found the central velocity of Hyades is at $\rm U=-40$\,\kms\, and that of Pleiades is at $\rm U=-25$\,\kms. \citet{Antoja12} measured the central values as $-30$\,\kms\, and $-16$\,\kms, respectively, in the U component with RAVE data. In any case the two over-densities are separated by 15\,\kms, larger than the uncertainty of the velocity in this work by a factor of 2. The validation test discussed in section~\ref{sect:deconv} demonstrates that our method is capable of distinguishing substructures with such separation in the U-V distribution. Therefore, the merging of Hyades and Pleiades over-densities in the LAMOST pilot survey data may either be because of fluctuations within a larger substructure, or due to different sampling volumes with \textit{Hipparcos} and RAVE. 

Although the Hercules stream (structure 7) is also not separated from other structures, it shows a clear asymmetry in Figure~\ref{fig:uv}. Recall that we select only the F and G dwarf stars (relatively young) in our samples. The non-separation of the Hercules stream is consistent with \cite{Dehnen98}, in which the author showed that the Hercules stream is not as prominent for stars with $0<B-V<0.6$ than older stars with $B-V>0.6$. This implies that it is composed of relatively old populations, although the age range may be very broad, according to \citet{Famaey05}.  Compared with \citet{Antoja12}, which fixed the Hercules stream at $\rm V\sim-50$\,\kms, the LAMOST data has a slightly faster V velocity at about $-40$\,\kms. 

Another hot over-density is structure 8, which is located at $\rm V\sim-70$\,\kms\ and extends by around 100\,\kms\ in U. This is consistent with the over-density discovered by \citet{Arifyanto06}. Indeed, because the central position of the over-density is around $\rm U\sim50$\,\kms, we can obtain that $\sqrt{U^2+2V^2}\sim111$\,\kms, which is completely in agreement with their study. 

The Arcturus group (\citealt{Eggen71}; marked as structure 9) is also seen at $\rm V\sim-100$\,\kms\ close to the bottom of Figure~\ref{fig:uv}. Here, with a larger sample, we confirm that the Arcturus group is also elongated along $U$, consistent with groups 14, 17, and 19 in \citet{Antoja12}. The most over-dense region of the elongated substructure is at $\rm U\sim-25$\,\kms, which overlaps well with their group 14.

Combining the images of the two elongated hot substructures at $\rm V\sim-70$ and $-$100\,\kms, it seems that they form a wave-like picture. \citet{Minchev09} argued that such structures may arise from the unrelaxed disk $\sim1.9$\,Gyr ago due to the perturbation of the Galactic bar formation.

\subsection{New substructures}\label{sect:newsub}

A few new substructures are unveiled in the U-V distribution. 
The most obvious substructure is the thin ripple-like structure 1 at the top of the majority of the stars found in Figure~\ref{fig:uv}. Because we use multiple Gaussians to build the empirical model of the U-V distribution and a Gaussian profile can not be made to form a ripple-like structure, the algorithm automatically selects two Gaussians with elongated covariance matrices to reconstruct such a substructure following the observed data. The substructure is also seen, somewhat blurred, in the top panel of Figure~\ref{fig:meanuv} in which the PDF of the distance is replaced with the mean distance.  This elongated feature is not prominent in other survey data, e.g. \textit{Hipparcos} (\citealt{Dehnen98, Bovy09} etc.), GCS \citep{Zhao09}, and RAVE \citep{Antoja12, Antoja13}. This is probably due to two reasons. Firstly, the \textit{Hipparcos} and GCS only cover stars within 100\,pc around the Sun, while the LAMOST pilot data can reach as far as 500\,pc. Although the RAVE data can reach a similar depth, it covers only the Southern Galactic cap, and with only a small fraction of the sky overlapping with the LAMOST survey. Secondly, some previous works used the wavelet transform method, which concentrates more on the small scale over-densities, and thus probably filters out such larger scale features.

Structure 1 has a narrower dispersion in V with an extended dispersion in $U$, which leads to a significantly larger $\sigma_{\rm U}/\sigma_{\rm V}$ ($\sim3$, approximated from the ratio of the major- and minor-axis of the contours in Figure~\ref{fig:uv}) than that for the whole sample ($<2$). It can not be explained by the population being kinematically hot, but may be associated with a resonance, which enhances only the radial excursion of the stellar orbits but not the azimuthal velocity. 
Meanwhile, the relatively higher V of the substructure indicates that many stars move faster than the local circular speed. Hence, their orbital guiding centre radii may be beyond the solar circle. In other words, this population should be from the outer disk. If the resonance is induced by the central rotating bar, then according to equations (3.150) and (3.80) in \citet{Binney08}, the 1:1 OLR is indeed around $R=10.6$\,kpc if the circular speed is 220\,\kms\ and the bar pattern speed is 50\,\kms\ kpc$^{-1}$. 
The complete orbital phases of stars experiencing the resonance of the Galactic bar should form a narrow annulus with somewhat large radius in the U-V plane. For the case of 1:1 OLR, the resonance radius should be beyond the solar orbit, and therefore, the 1:1 OLR Êstars observed in the solar neighborhood are only samples with orbital phase close to the pericenter. This leads to an incomplete annulus at around the maximum of V in the U-V plane, similar to feature 1.
However, the pattern speed can be as low as 40\,\kms\ kpc$^{-1}$ \citep{Long13} and the local circular speed as high as 250\,\kms\ \citep{Reid09}. Therefore, the position of the OLR is uncertain and the precise nature of the resonance is not conclusive. For instance, if we adopt the pattern speed of the bar inferred from \citet{Long13}, and the local circular speed of 220\,\kms, the 2:1 outer Lindblad resonance is then at $R=9.4$\,kpc, which is more likely to be responsible for the ripple-like structure.   

Just above structure 1, the new ridge-like structure 2 is located from $(-15, 15)$ to $(-60, 40)$ \kms. A similar feature also appears in the GI bottom left panel of Figure 3 in \citet{Dehnen98}, in which a ridge-like structure with (U, V) from $(-5, 10)$ to $(-30, 40)$ \kms\ is indicated. The B2, B3, B4 and AL panels of Figure 3 of Dehnen (1998) also show some evidence of feature 2. It is not clear if this feature is an extension of the Sirius over-density (labelled as 3 in Figure~\ref{fig:uv}) since they are apparently connected with each other. It seems that member stars moving out with higher speed (i.e., smaller negative U) have larger angular momenta (i.e., larger V). 

Yet another new over-density, structure 6, is found at $(-102, -24)$\,\kms. It may be either a new feature or the tail of the Hercules stream containing a clump of stars moving outward.
  
\section{The U-V distribution at different Galacto-centric radii}\label{sect:discussions}

The substructure induced by the resonance of the bar and spiral arms in the velocity distribution may vary with position. \citet{Dehnen00}  mapped various simulated U-V distributions at different radii and azimuthal angles with respect to the central bar. With the RAVE data, \citet{Antoja12, Antoja13} found that, indeed, the location of the Hercules stream and the gap between the Hercules stream and the majority of the stars varies with radius. The variation of the Hercules stream at different locations may well constrain the pattern speed and azimuthal angle of the Galactic bar.
The volume of the LAMOST pilot data used in this work is similar to that of RAVE. Therefore, it is worth investigating the variation of the U-V distribution for stars inside of, and outside of the solar circle ($R_{\rm 0}\simeq 8$ kpc).

Although using the PDF of the distance with Monte Carlo simulation can gain a slightly better resolution of the substructures, it is very time consuming. Hence, we use the median distance estimated from the PDF of distance to map the U-V distribution for two subsamples with $R<R_{\rm 0}$ and $R>R_{\rm 0}$, respectively. In these fits, we use 31 Gaussians and the regularization parameter, $w$, is taken to be 64 \kmss.

We first verify if the median distance can give the same U-V distribution as that of the PDF of the distance. The top panel of Figure~\ref{fig:meanuv} shows the U-V distribution for the whole sample with median distance.  The substructures appear much broader than those in Figure~\ref{fig:uv}, because (1) there are far more Gaussians used in Figure~\ref{fig:uv} than in the top panel of Figure~\ref{fig:meanuv}; and (2) since the uncertainties of U and V are slightly larger due to the additional contribution of the uncertainty in the distance, we set w=64 \kmss\ in Figure~\ref{fig:meanuv}. Even so, most of features except structure 9 are still recognizable.
Therefore, we can use the median distance to map the U-V distribution for the stars located inside and outside the solar circle. The other panels of Figure~\ref{fig:meanuv} show the two U-V distributions with $R<R_{\rm 0}$ (middle panel) and $R>R_{\rm 0}$ (bottom panel).

The Hyades-Plaiedes over-density (structure 5) is still seen in the U-V distribution inside the solar circle but is no longer recognizable in the one outside the solar circle. This implies that most of its member stars are located inside the solar circle. Moreover, below the Hyades-Pleiades structure, there is an extended feature from $(-10, -15)$ to $(5, -50)$ \kms\ in the middle panel of Figure~\ref{fig:meanuv}. This feature can also be identified in the third column panel of Figure 3 of \citet{Antoja12}.

The Hercules stream (structure 7), on the other hand, is clearly separated from the majority of the stars in the U-V distribution outside the solar circle, but not separable in the U-V distribution inside the solar circle. This is in agreement with \citet{Antoja12, Antoja13}.

In the U-V distribution outside the solar circle, structure 10 (which should be the Wolf 630 over-density according to \citet{Antoja12}) is clearly seen just below the Sirius over-density, while it is not recognizable in the U-V distribution inside the solar circle. This is also approximately consistent with the RAVE results demonstrated by \citet{Antoja12}.

The ripple-like structure 1, which is prominent in Figure~\ref{fig:uv}, is no longer clearly seen in the U-V distribution neither inside nor outside the solar circle. However, structure 2 is even more prominent in the U-V distribution inside the solar circle.

It is not easy to find the origins of the variations of the substructures with Galacto-centric radii. \citet{Quillen11} showed a possible scenario from their N-body simulation where these subtle substructures may be associated with the relative locations of the spiral arms.  \citet{Antoja11} inferred that the inner Lindblad resonances induced by the spiral arms at different radii may also produce different patterns of substructures in the U-V distribution.

\section{Conclusions}\label{sect:conclusions}

The goal of the LAMOST pilot survey is to verify the survey design and to test the performance of the instruments: it is not expected to be as good as the formal (ongoing) survey. Even so, the pilot survey has already collected sufficient data for an investigation of the local velocity distribution. Radial velocity estimation is sufficiently accurate to distinguish kinematic substructures with velocities around 6-10\,\kms. With these data, we are able to identify known and new substructures in the U-V velocity distribution. Three new substructures, 1, 2, and 6 (see Table 1 and Figure~\ref{fig:uv}), are found from the data. Structure 1 is likely associated with the resonance induced by the central Galactic bar. Structures 8  and 9 (Acturus group) are consistent with a scenario where the local disk is being perturbed by the Galactic bar according to \citet{Minchev09}. 

When we separate the data into two samples at the solar circle, the U-V distributions of the two groups of stars are significantly different. The Hercules stream is more isolated in the sample beyond the solar circle, while the Hyades-Pleiades substructure is more prominent in the sample inside the solar circle. The latter could be associated with the spiral structures of the Milky Way, but it remains a puzzle how the spiral structures can produce such spatial variations. More data covering larger ranges of distances is required to constrain better the origin of these subtle substructures on nearly circular orbits. We plan to investigate these issues with the LAMOST DR1 data in a future study.

\section{Acknowledgment}
We thank the referee for his (or her) detailed comments that have helped improve this paper substantially.
This work is supported by the National Science Foundation of China under grant No. 11373032, 11333003 (CL and SM) and 11390372 (SM). 
CL acknowledges the Major State Basic Research Development Program  2014CB845704. 
This work has also been supported by the Strategic Priority Research Program ``The Emergence of Cosmological Structures'' of the Chinese Academy of Sciences Grant No. XDB09000000 (SM and CL).
The Guoshoujing Telescope (the Large Sky Area Multi-Object Fiber Spectroscopic Telescope LAMOST) is a National Major Scientific Project built by the Chinese Academy of Sciences. Funding for the project has been provided by the National Development and Reform Commission. LAMOST is operated and managed by the National Astronomical Observatories, Chinese Academy of Sciences.

\appendix

\section{Proper motion correction}\label{sect:pm}
The proper motions in the PPMXL catalogue \citep{Roser10} have apparent systematic errors which will affect the U-V velocity distributions. If we assume the systematic error is 1 $\rm{mas\ yr^{-1}}$ and the distance to a star is 500 pc, the difference in U-V velocity plane will be about 2 \kms. On the other hand, the mean errors of the proper motions in the PPMXL Catalogue vary from 4 $\rm{mas\ yr^{-1}}$ to more than 10 $\rm{mas\ yr^{-1}}$, depending on stellar magnitudes. That is, the errors of many stars in our F\&G sample are more than 10 \kms\ which will smooth out the results due to the EM method. We need therefore to minimize the errors in the proper motions.

Quasars are very distant objects whose proper motions are essentially zero.  The sample has 151,107 quasars from the cross identification between PPMXL and SDSS. The systematic deviations from zero of the proper motions of the quasars represent the systematic errors in the proper motions of the stars, and their standard deviation represents the random error \citep{Wu11}. The error of the stellar proper motions can be replaced by the random error.

\begin{figure}
\centering
 \includegraphics[scale=0.3]{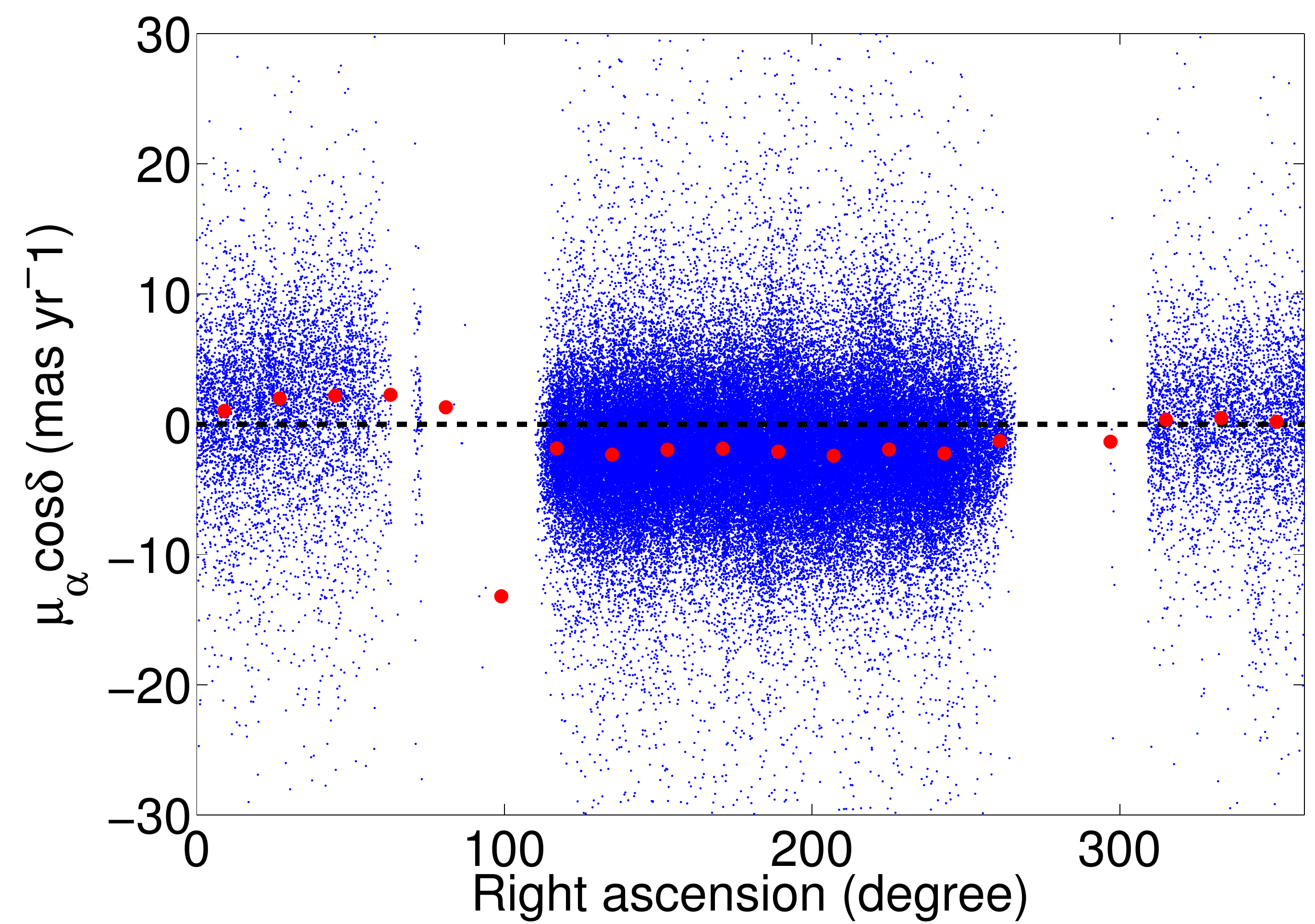} 
 \caption{The proper motions of QSOs in the PPXML catalogue. The black dashed line indicates zero systematics. The red points show the median values in each bin of right ascension.}\label{fig:qsos}
\end{figure}

Figure~\ref{fig:qsos} shows how the proper motions of quasars vary with the lines of sight which reflects the systematic errors of PPMXL. The blue points are for the quasars, the red points are the mean proper motion of the quasars in different right ascensions, and the black dashed line indicates zero systematic bias. The systematic errors change with position and the values are about 2 $\rm{mas\ yr^{-1}}$, consistent with \citet{Wu11}.

Since the systematic errors are correlated with position, we need to correct the proper motions of stars. For each star, first, we find the quasars within a circle of diameter 2 degrees. Then, as a subsample, we calculate their median value and dispersion. The median value is used to correct thet proper motion, and the dispersion is used as the error of the proper motion. The errors are typically around 4 $\rm{mas\ yr^{-1}}$.

\section{The choice of the number of Gaussians in the reconstructed U-V distribution}

Two parameters determine the reconstructed U-V distribution assembled by multiple Gaussians, namely the regularization parameter, $w$, and the number of Gaussians, K. The former has been discussed in section~\ref{sect:deconv}, in this section we discuss how to determine the latter, K.
In principle, the larger K used in the model, the better it can fit the data. In practice, too many Gaussians increase the complexity and some may follow the statistical fluctuations, leading to overfitting. Therefore, we need to determine how many Gaussians are suitable for our mixed model.

There are multiple means to provide statistically optimized choice of K according to \citet{Bovy09}. These are split out into internal and external means. The internal means provide certain criteria that can be applied to determine the number of K, e.g., Akaike's information criterion \citep[]{Akaike74}, minimum description length \citep[]{Rissanen78,Schwarz78}, minimum message length \citep[]{Wallace87,Oliver94,Oliver96},   and Bayesian evidence \citep{Roberts98}, etc.. However, these criteria usually do not agree with each other. Moreover, Bovy et al. discussed that the unknown covariance in the data may also affect the determination of K.
Another kind of means are the external validation. Since the \textit{Hipparcos} data used by Bovy et al. generally do not have radial velocities, they used the radial velocities provided by GCS catalog as the external source to validate the velocity distribution model and find the most appropriate K. Because it is one of the most conservative solutions, it seems that the validation test prefers a smaller number of Gaussians in \citet{Bovy09} and, consequently, it may miss some substructures because of the lack of Gaussians to model them. Moreover, it also relies on the extra information in a subsample, in the case of the data used by \citet{Bovy09},  it relies on the radial velocity of the GCS data. In general, it is difficult to define such a subsample without systematic bias. For our data, since we use all three dimensional velocities, there is no such extra information which can be applied to validate the best choice for K.

\begin{figure*}
\centering
\begin{minipage}{17.5cm}\begin{center}
 \includegraphics[scale=0.3]{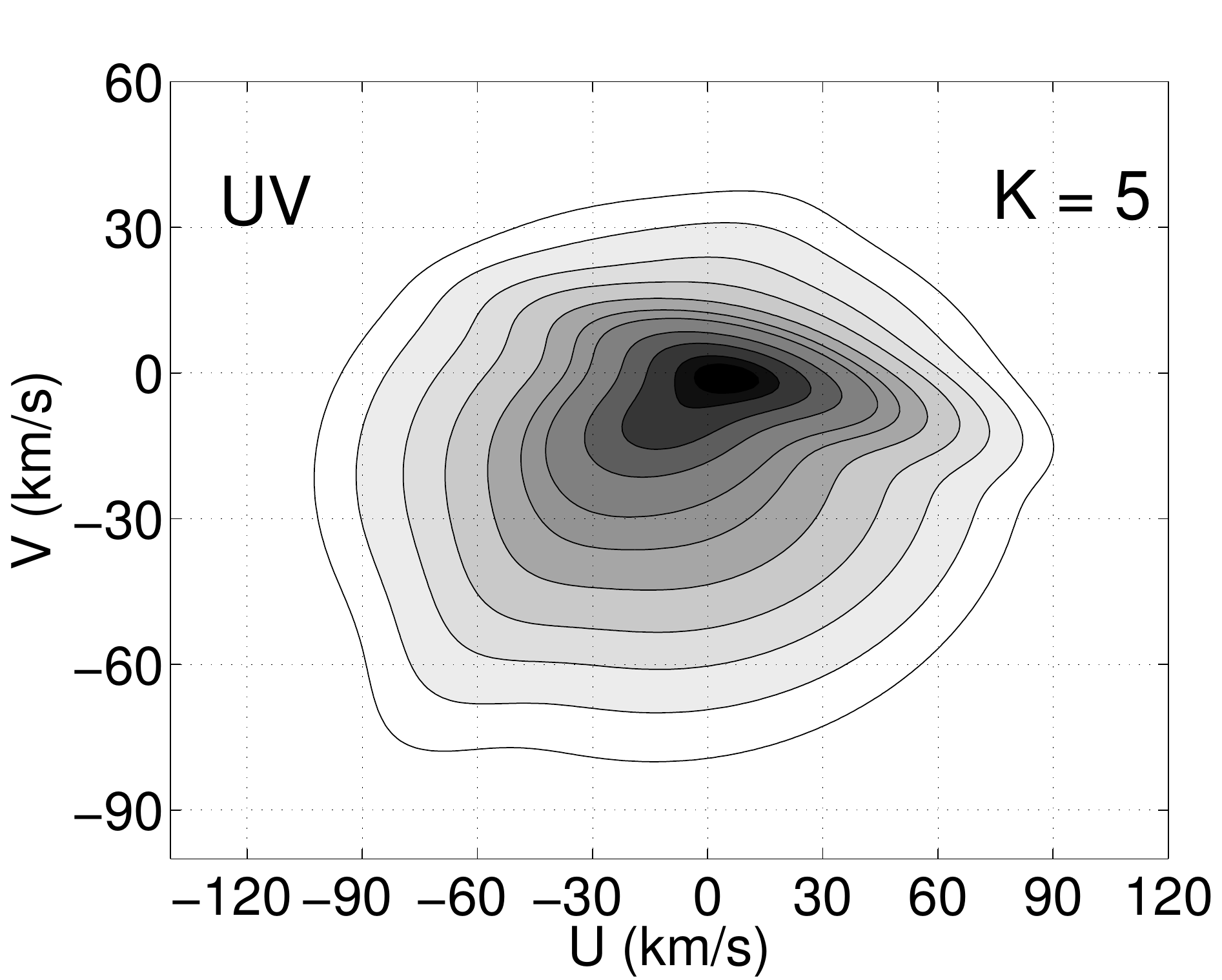}
 \includegraphics[scale=0.3]{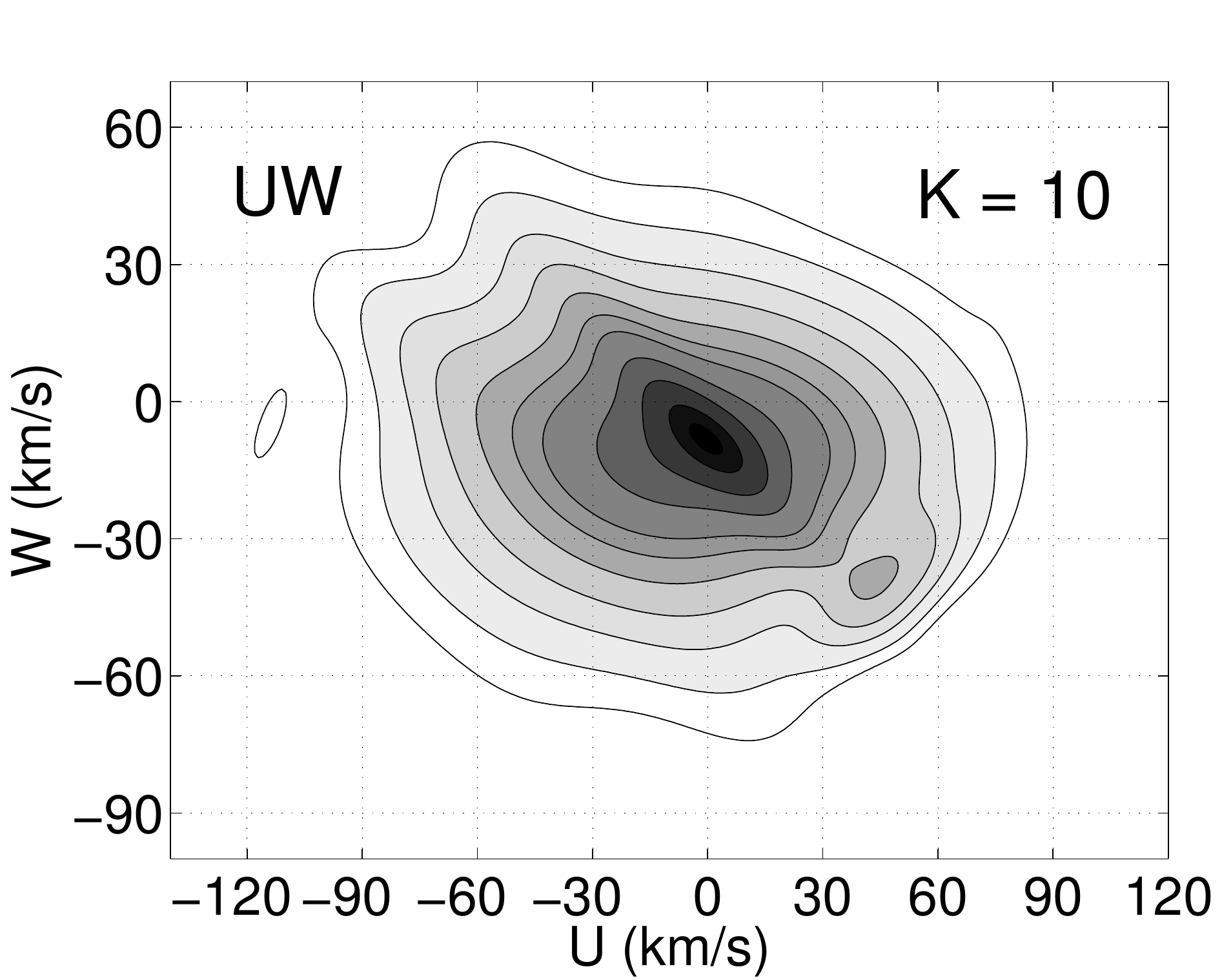}
 \includegraphics[scale=0.3]{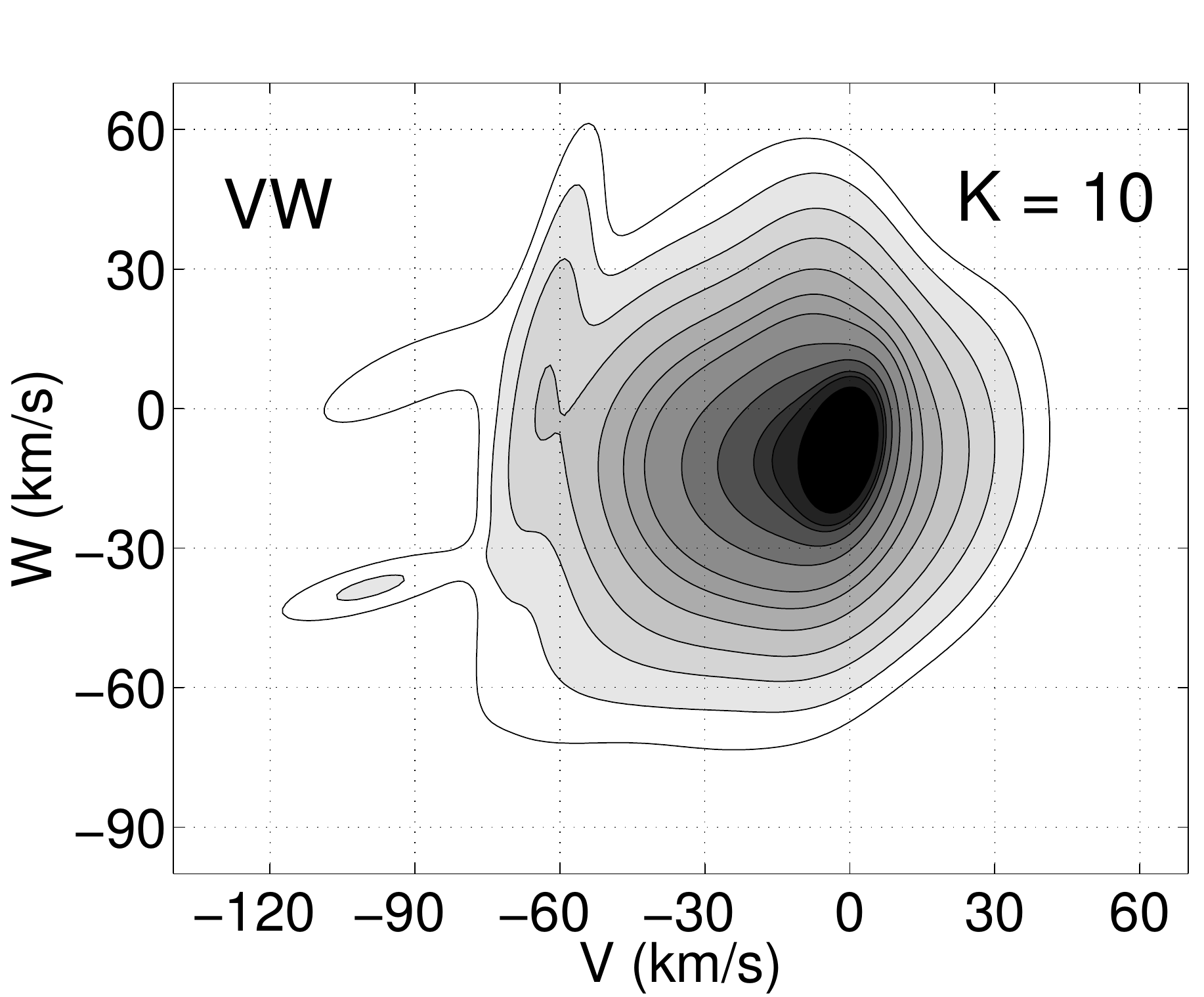} 
 \includegraphics[scale=0.3]{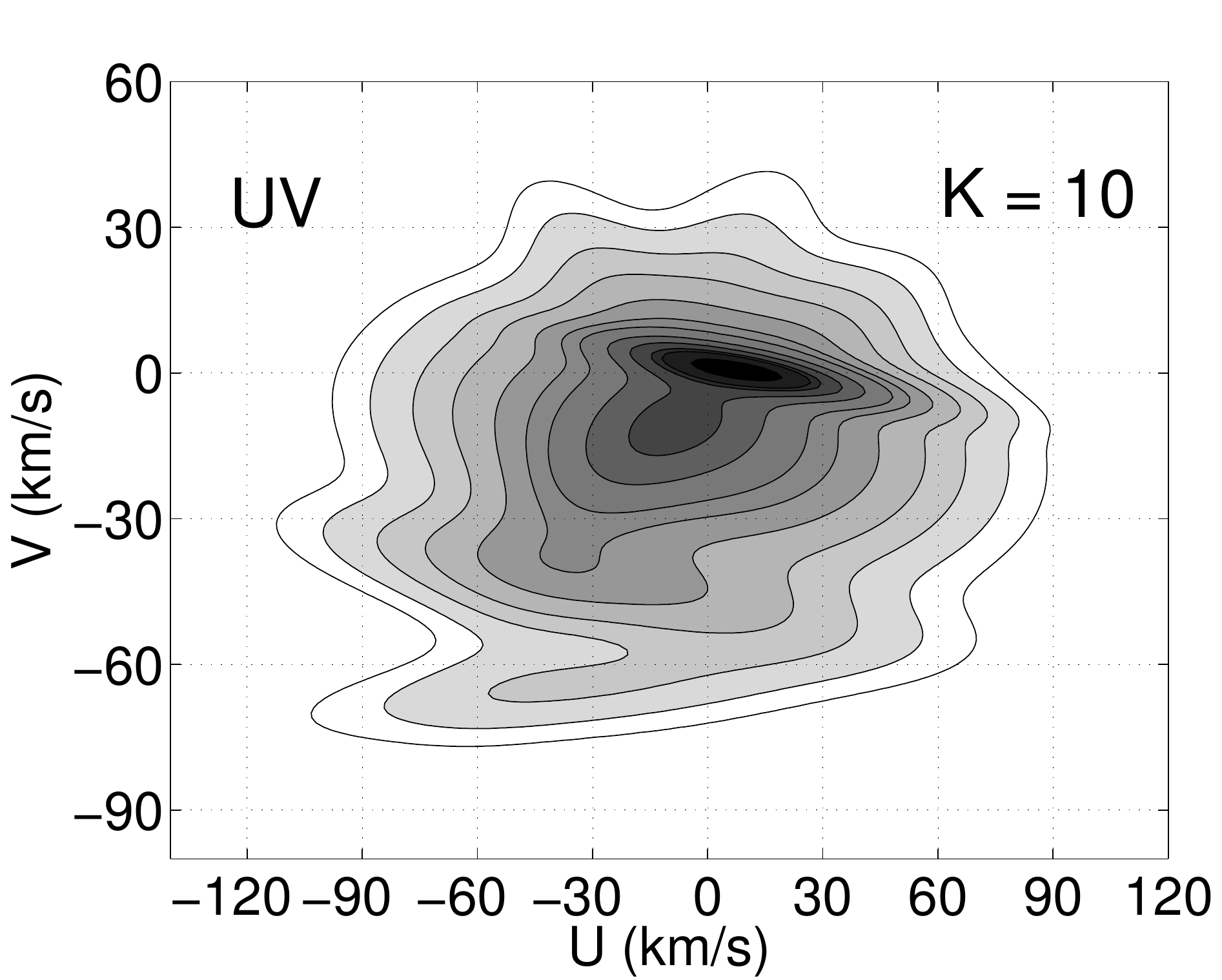} 
 \includegraphics[scale=0.3]{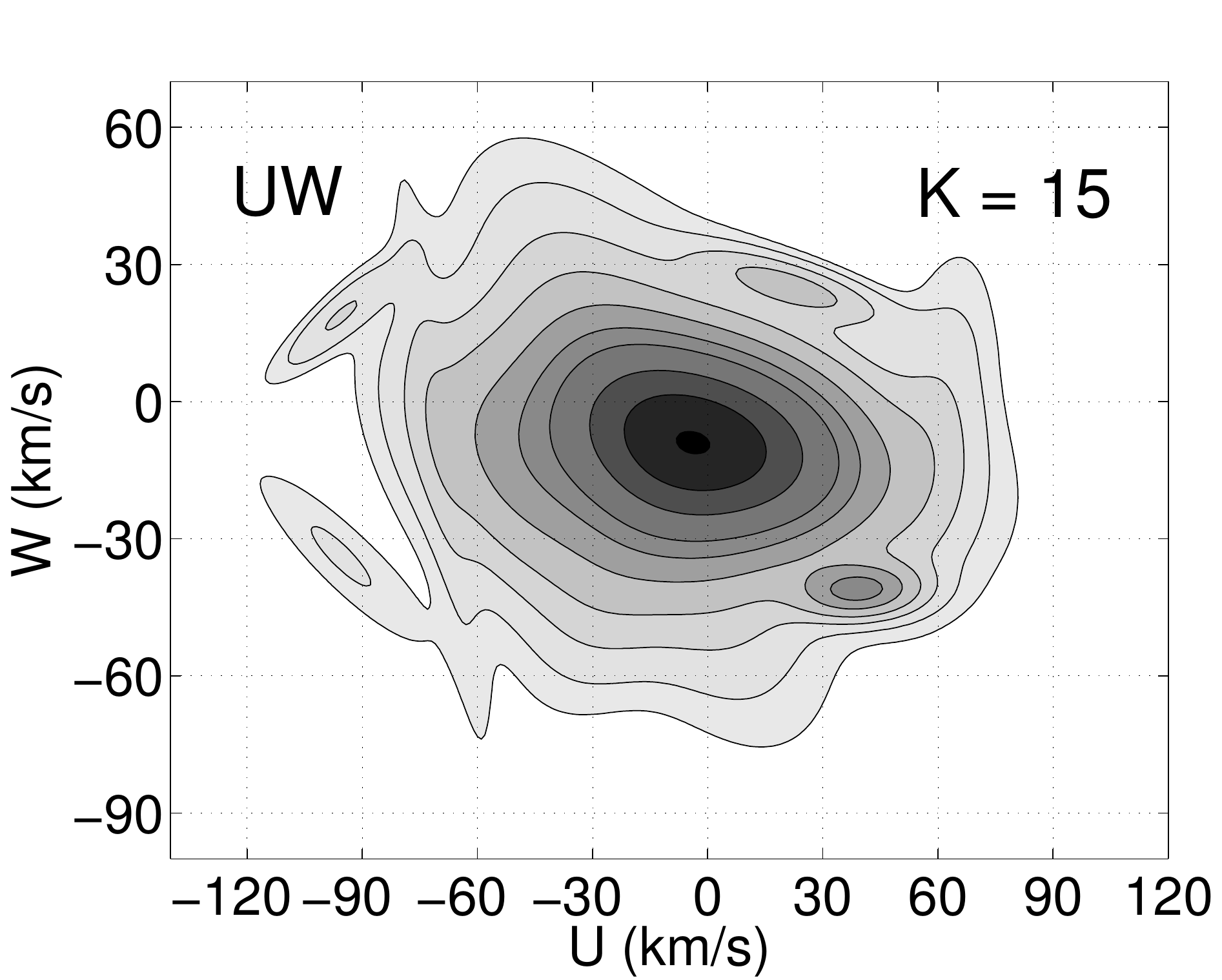} 
 \includegraphics[scale=0.3]{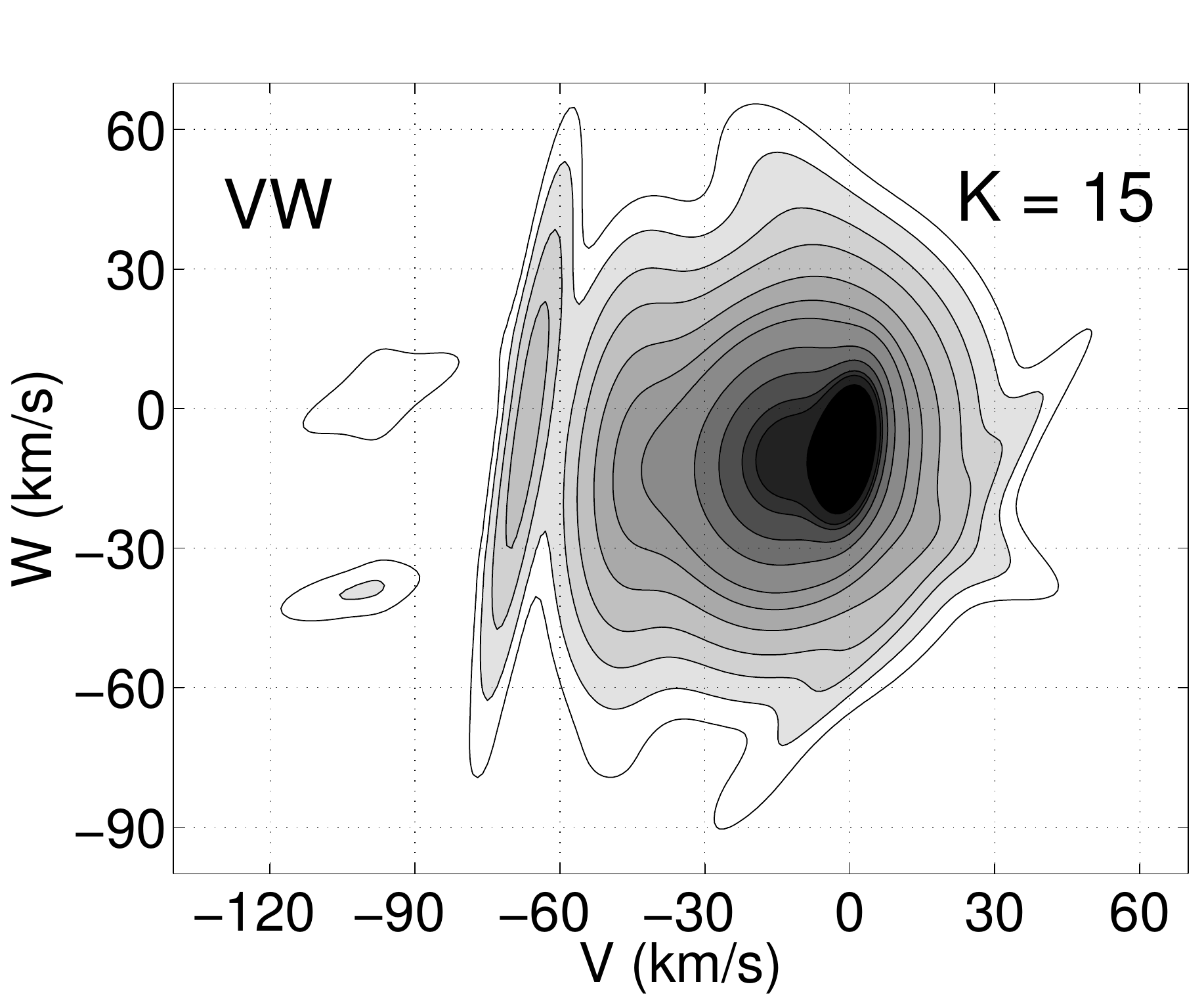} 
 \includegraphics[scale=0.3]{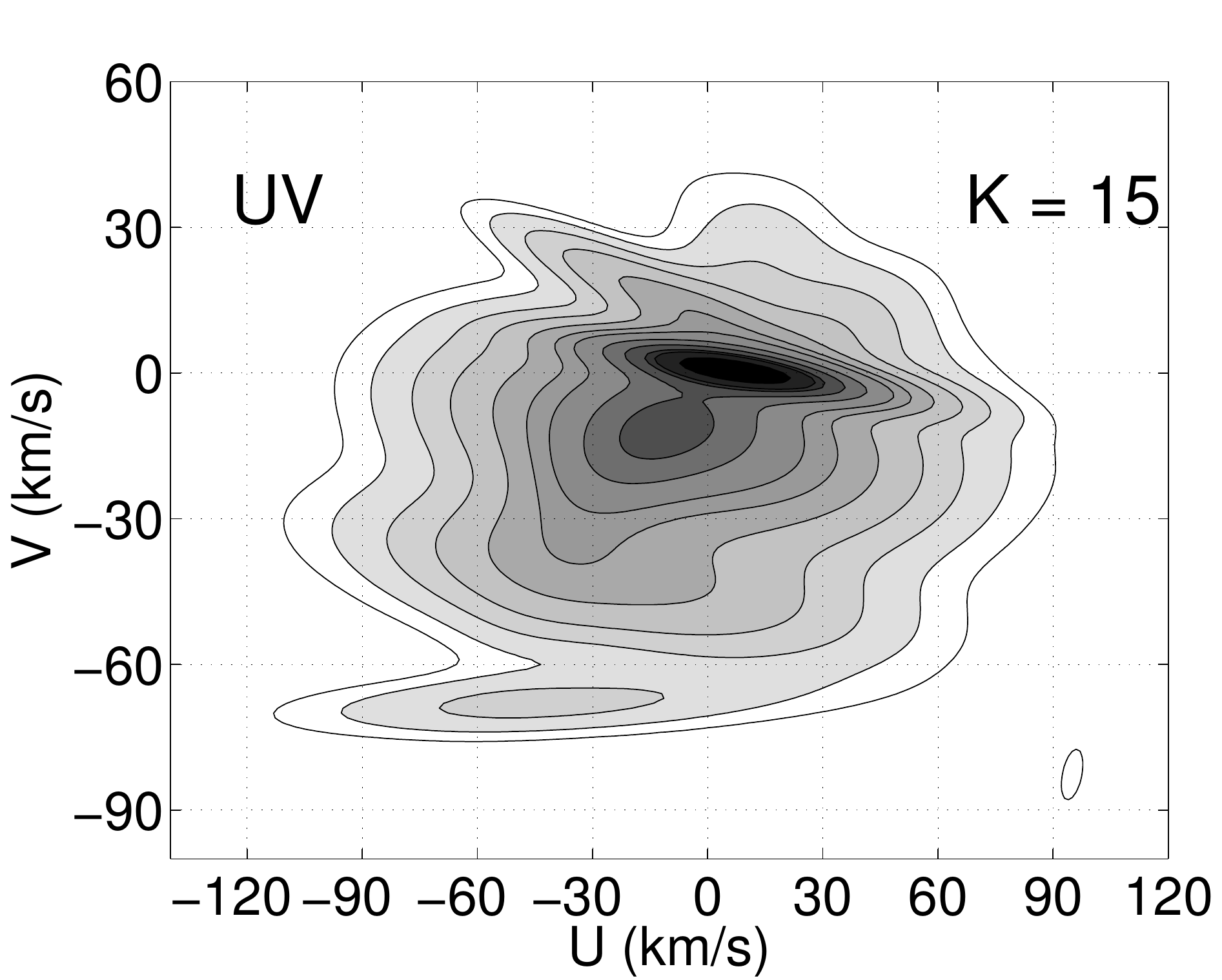} 
 \includegraphics[scale=0.3]{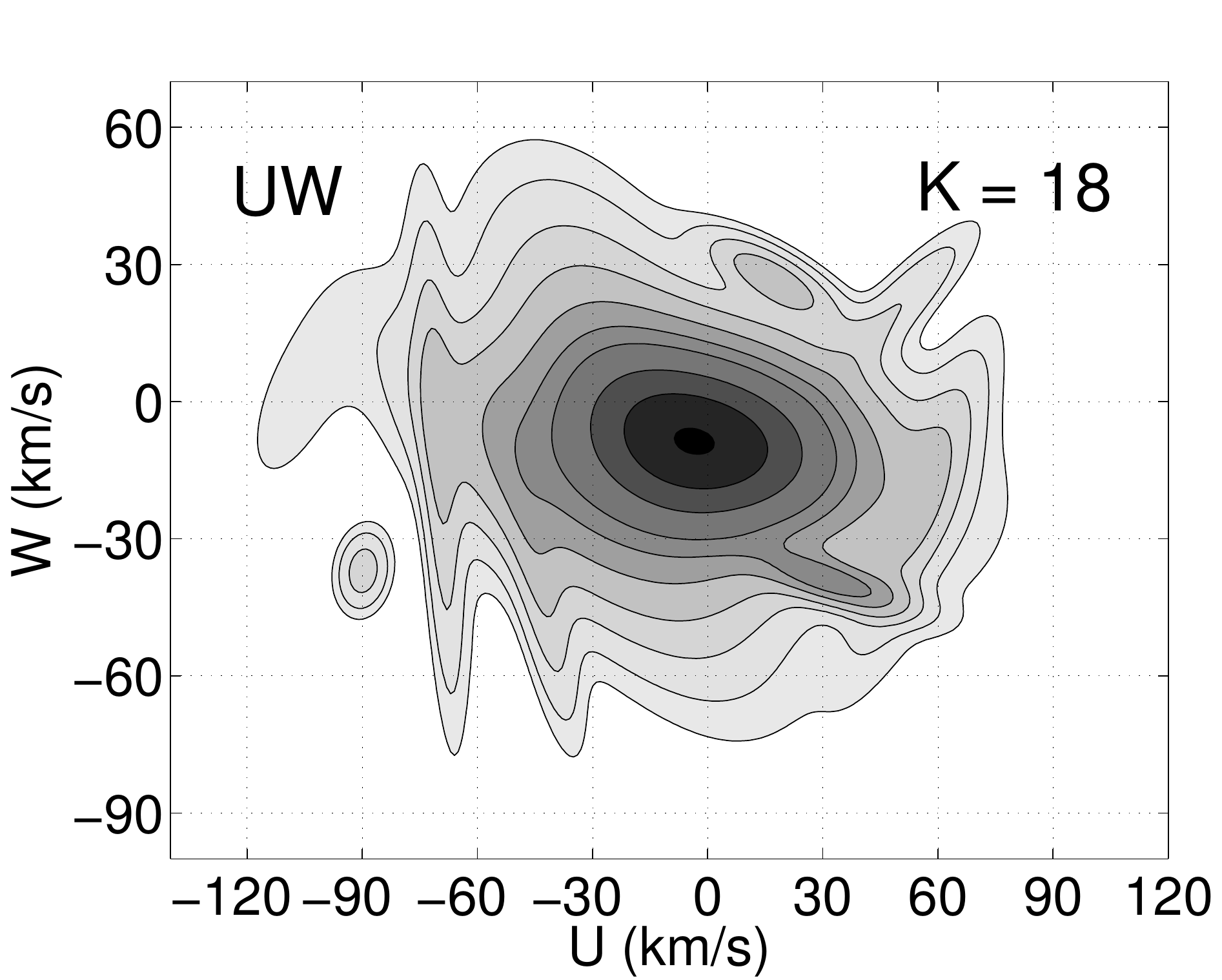} 
 \includegraphics[scale=0.3]{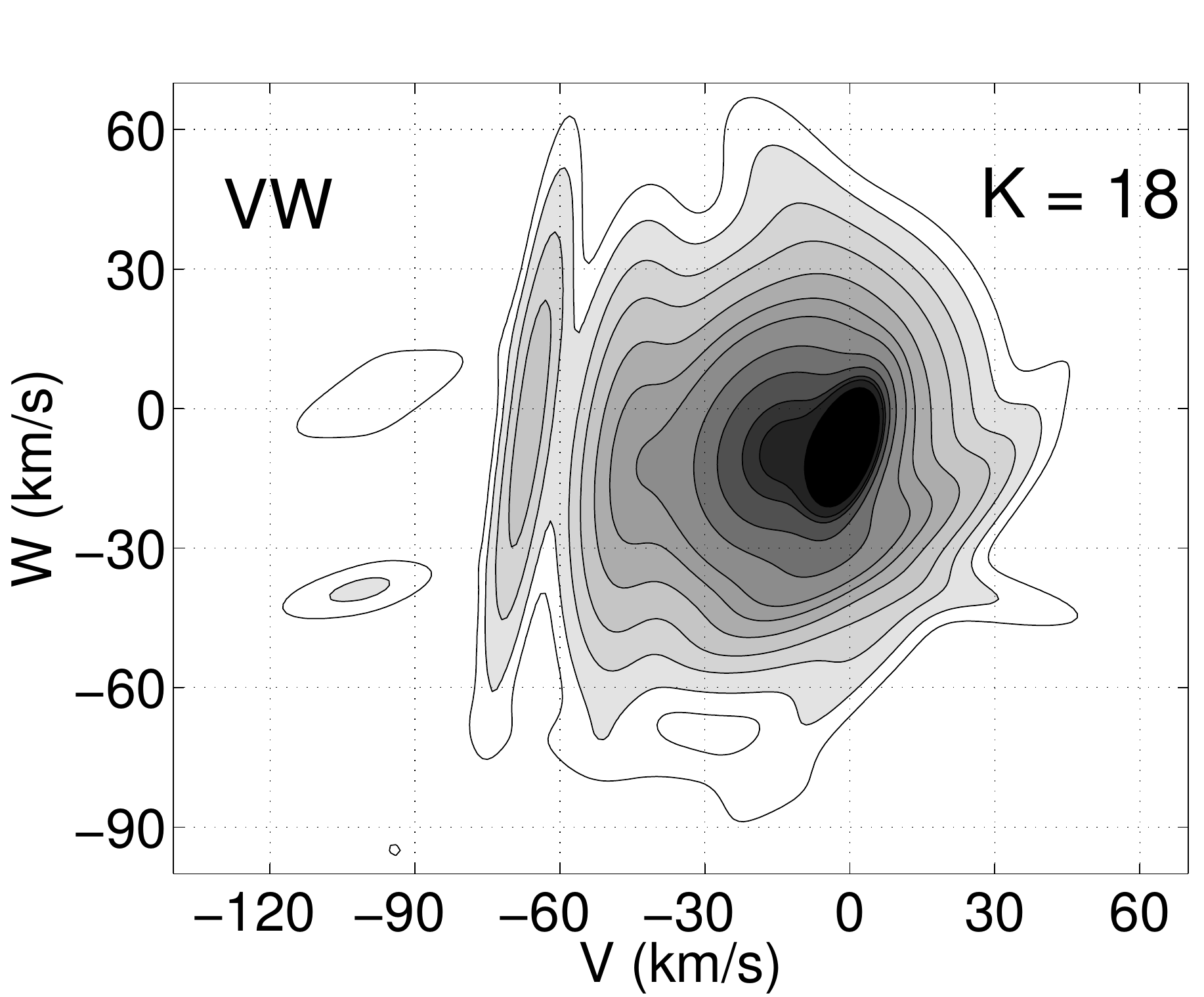} 
 \caption{In the left column, the top, middle and bottom panel shows the U-V distribution with K=5, 10 and 15 Gaussians respectively. In the middle column, the top, middle and bottom panel shows the U-W distribution with K=10, 15 and 18 Gaussians. In the right column, the top, middle and bottom panel shows the V-W distribution with K=10, 15 and 18 Gaussians. The contour levels are the same as in Figure~\ref{fig:uv}. All the results are fitted using the mean distances.}\label{fig:UVK5}
\end{center}
\end{minipage}
\end{figure*}

Thus, we turn to another experimental means to determine the value of K for our case. We reconstruct the U-V distribution based on the average distances with K=5, 10, and 15 Gaussians (see the left column of Figure~\ref{fig:UVK5}) and check if the most prominent substructures shown in Figure ~\ref{fig:meanuv} also appear when the model contains fewer Gaussians. The left column of Figure~\ref{fig:UVK5} shows that the most interesting features, e.g., right part of 1, 2 and 8, indeed show up even with K=5. This implies that these prominent features  cannot be spurious results due to the use of too many Gaussians.

Alternatively, testing the smoothness of U-W and V-W distribution can also qualitatively investigate whether the choice of K is suitable, given the prior knowledge that the velocity U-W and V-W distributions should be smoother than that in the U-V plane. In the middle and right columns of Figures~\ref{fig:UVK5}, we show the U-W and V-W distributions with K=10, 15, and 18 Gaussians, respectively. They show that spurious structures increase with K but do not diffuse to the central region of the U-W and V-W distributions, i.e., ~50 km/s around the center, but only contribute to the outskirts, which is mostly due to relatively sparse data there. It is notable that two structures located at (U, W) = (30, -40) and (U, W) = (27, 28) do exist in the U-W distribution when K=15 and 18, which may indicate two real structures. Because most of the data concentrate in the central region (a 10 \kms\ $\times$ 10 \kms\ bin), the signal-to-noise ratio (S/N) of the stellar density in the center is about 20 according to the Poisson distribution. Hence, statistical fluctuations in the region with such high S/N should be weak and the Gaussians assigned in the central region are unlikely to be affected by the noise. Although W distribution is smoother, the U-W and V-W distributions may still have few structures, such features have also been seen in Figures 3 and 4 of \citet{Bovy09}. In the outskirts, on the contrary, since the density is low, the S/N of the stellar density is also low and the fluctuation in this region is mainly arbitrary. When we add more Gaussians in the U-W and V-W distributions, since the central parts are quite smooth and do not need many Gaussians to be fit, the additional Gaussians are assigned to the outskirts to fit the statistical fluctuations and create possible, spurious structures. That is, spurious structures may firstly occur in the region with lowest S/N of the stellar density; only when the low S/N region has been well covered and if there are still a few Gaussians left, they will tend to overfit the weak arbitrary fluctuations in the high S/N region. 

For the case of the U-V distribution, since the central region (with high S/N of the stellar density) does have some substructures, it needs more Gaussians to fit these features and thus not many Gaussians are left to overfit the outskirts. Therefore, we do not see strong spurious structures in the top panel of Figure~\ref{fig:meanuv}, implying that the choice of K is suitable for our data. However, when we look at the middle and bottom panels of Figure~\ref{fig:meanuv}, because these are the U-V distributions for two subsamples located within and outside the solar circle, respectively, some Gaussians are assigned to their outskirts with lower S/N. Subsequently, these two panels  show more spurious structures than the top panel.

Combining these two independent means, we can infer that K=20 will not produce artificial spurious structures in the central region of the U-V distribution. Given that the prominent substructures shown in the top panel of Figure 8 have already shown their significance when K is smaller, we believe they are real features.

\label{lastpage}
\end{document}